\documentclass[twocolumn]{aastex631}

\usepackage{graphicx}	
\usepackage{amsmath}	
\usepackage{subfigure}
\usepackage{natbib}
\usepackage{color}
\usepackage{tabularx}
\usepackage{longtable}
\usepackage{bm}
\usepackage{threeparttable}
\usepackage{enumerate}
\usepackage[figuresright]{rotating}
\usepackage[normalem]{ulem}
\usepackage{verbatim}
\usepackage{diagbox}
\usepackage{slashbox}

\newcommand{\code}[1]{{\texttt{#1}}}

\newcommand{\kepler}{{\it Kepler}}

\begin{document}


\title{Metallicity Dependence of Giant Planets around M Dwarfs}

\correspondingauthor{Tianjun Gan}
\email{tianjungan@gmail.com}

\author[0000-0002-4503-9705]{Tianjun~Gan}
\affil{Department of Astronomy, Tsinghua University, Beijing 100084, People's Republic of China}

\author[0000-0002-9807-5435]{Christopher A. Theissen}
\affil{Center for Astrophysics and Space Sciences, University of California, San Diego, 9500 Gilman Dr, La Jolla, CA 92093, USA}

\author[0000-0002-6937-9034]{Sharon X. Wang}
\affil{Department of Astronomy, Tsinghua University, Beijing 100084, People's Republic of China}

\author[0000-0002-6523-9536]{Adam J.~Burgasser}
\affil{Center for Astrophysics and Space Sciences, University of California, San Diego, 9500 Gilman Dr, La Jolla, CA 92093, USA}

\author[0000-0001-8317-2788]{Shude Mao}
\affil{Department of Astronomy, Tsinghua University, Beijing 100084, People's Republic of China}



\begin{abstract}

We investigate the stellar metallicity ([Fe/H] and [M/H]) dependence of giant planets around M dwarfs by comparing the metallicity distribution of 746 field M dwarfs without known giant planets with a sample of 22 M dwarfs hosting confirmed giant planets. All metallicity measurements are homogeneously obtained through the same methodology based on the near-infrared spectra collected with a single instrument SpeX mounted on the NASA Infrared Telescope Facility. We find that 1) giant planets favor metal-rich M dwarfs at a 4-5$\sigma$ confidence level, depending on the band of spectra used to derive metallicity; 2) hot ($a/R_\ast\leq 20$) and warm ($a/R_\ast> 20$) Jupiters do not show a significant difference in the metallicity distribution. Our results suggest that giant planets around M and FGK stars, which are already known to prefer metal-rich hosts, probably have a similar formation channel. In particular, hot and warm Jupiters around M dwarfs may have the same origin as they have indistinguishable metallicity distributions. With the refined stellar and planetary parameters, we examine the stellar metallicities and the masses of giant planets where we find no significant correlation. M dwarfs with multiple giant planets or with a single giant planet have similar stellar metallicities.  Mid-to-late type M stars hosting gas giants do not show an apparent preference to higher metallicities compared with those early-M dwarfs with gas giants and field M dwarfs. 


\end{abstract}



\section{Introduction} \label{sec:intro}

Since the first detection of a hot Jupiter around a solar-type star outside the Solar System \citep{Mayor1995}, the formation channel of giant planets has been under debate over the last two decades. One of the leading hypotheses is core accretion \citep[e.g.,][]{Pollack1996,Ida2004coreaccretion,Mordasini2008}, a bottom-up mechanism starting with a massive solid core that eventually grows into a giant planet through runaway gas accretion before the disk dissipates. 

Several pieces of observational evidence have been put forward in support of such a scenario. For example, metal-rich stars appear to be more likely to host cold giant planets found by radial velocity (RV) surveys \citep[e.g,][]{Gonzalez1997,Santos2004,Fischer2005,Johnson2010,Montet2014,Adibekyan2019}, the so-called giant planet-metallicity correlation. Under the core accretion paradigm, as the bulk metallicity is supposed to reflect the amount of materials available in the protoplanetary disk, one would naturally expect that gas giants form more easily around stars with higher metallicity. \cite{Maldonado2012} went a step further and compared the stellar metallicity preference of hot and cool Jupiter systems, where the authors pointed out a lower frequency of hot Jupiters than cool ones at low metallicities. Later work from \cite{Petigura2018} looked into the metallicity dependence of planet occurrence rate with a sample of \kepler\ targets \citep{Borucki2010} and reported a steeper trend towards short-period gas giants. With a larger sample, \cite{Narang2018} studied the average metallicity of stars with planets in two period bins split at 10 days. Although stars with short-period small planets ($M_p\leq 50\ M_\oplus$) tend to be more metal-rich than those hosting longer period ones, such a trend is not significant when moving to the giant plant branch, which further complicates the picture. More recently, \cite{Osborn2020} revisited the planet–metallicity correlation for hot Jupiters by comparing the metallicity distribution of hot Jupiter hosts and that of a field star population simulated using the Besan\c{c}on Galaxy model \citep{Robin2003}. They confirmed that hot Jupiters prefer metal-rich stars but the correlation coefficient is roughly comparable with gas giant planets with longer periods, indicating a similar formation origin. 

However, most aforementioned efforts were made to FGK stars, equivalent relevant studies on giant planets around M dwarfs are still lacking so far even though M stars are the dominant stellar population in the solar neighborhood \citep{Henry2006,Reyle2021}. This is not only because of the difficulty in determining the M dwarf stellar properties but also due to the low occurrence rate of both hot and cold Jupiters around M dwarfs in contrast to other types of stars. \cite{Gan2023} found a frequency of $0.27\pm0.09$\% for hot Jupiters around early-type M dwarfs and the number slightly decreases to $0.137\pm0.097$\% around late-type M stars \citep{Bryant2023,Pass2023}. Regarding cold Jupiters, \cite{Johnson2010} estimated $0.03\pm0.02$ giant planets per M star within 2.5 AU (see also \citealt{Sabotta2021}). Nevertheless, a few such high-mass-ratio systems \citep[e.g.,][]{Morales2019,Gantoi4201,Hartman2024,Bryant2024,Stefansson2024,Hotnisky2024} were found to stretch the core accretion theory and may favor the gravitational instability model \citep{Boss2002} instead, making them particularly crucial to understand the giant planet formation. Plenty of techniques such as broadband photometry-based empirical relations \citep[e.g.,][]{Terrien2012,Mann2013,Mann2015,Mann2019,Newton2014,Newton2015} and detailed spectroscopic analysis \citep[e.g.,][]{Veyette2016,Marfil2021,Passegger2022,Bello2023,Jahandar2024} have been developed to determine the stellar parameters of M dwarfs, allowing for better characterizations for M dwarf planetary systems \citep{Gore2024}. 

Based on photometric calibrations, early work from \cite{Johnson2009} found that M dwarfs with planets tend to have metallicities in excess of field M stars in the solar neighborhood. \cite{RojasAyala2010} later reported that M dwarfs hosting Jovian-like planets detected by RV surveys have higher metallicities than those with Neptune- or Earth-size planets but the sample size is limited. Using a sample of three M dwarfs with RV-detected giant planets, \cite{Gaidos2014metallicity} found that they are not significantly more metal-rich than the average parent sample and the power-law index relating planet occurrence rate to stellar metallicity is consistent with Sun-like star counterparts within uncertainties. In terms of hot Jupiters, the Transiting Exoplanet Survey Satellite \citep[TESS;][]{Ricker2015} has been enlarging the number, where most of them were detected around high-metallicity M stars \citep[e.g.,][]{Gantoi530,Kanodia2022,Kagetani2023,Han2024}. In addition, \cite{Gantoi4201} claimed a possible trend that hot Jupiters prefer more metal-rich M stars than warm Jupiters based on the literature metallicity measurements but the heterogeneous instruments and methodologies result in systematic biases, preventing a firm conclusion.

In this manuscript, we compare the metallicity ([Fe/H] and [M/H]) distributions of 22 M dwarfs hosting confirmed giant planets (defined as planets with mass $0.2\ M_{\rm Jup} \leq M_p<13.6\ M_{\rm Jup}$) with a field M dwarf sample and investigate the difference between the hot and warm Jupiter groups. All metallicities are homogeneously measured using the near-infrared low-resolution spectra from a single instrument SpeX \citep{Rayner2003} mounted on the 3.2-m NASA Infrared Telescope Facility (IRTF), through the same method. Compared with previous works \citep{Johnson2009,RojasAyala2010,Gaidos2014}, our giant planet sample is about four times larger and includes several short-period gas giants detected by the TESS mission, which allows us to separate our sample into hot and warm Jupiters and study the metallicity preference of these two sub-groups. We refine the stellar and planet properties of our planet sample and explore potential correlations with stellar metallicity. The rest of the paper is structured as follows. We begin with sample construction in Section~\ref{sample_construction}. We summarize our SpeX observations in Section~\ref{observation}. Section~\ref{analysis} describes how we determine the metallicity and refine the stellar and planet parameters. In Section~\ref{discussion}, we conduct the metallicity comparison between two samples. We conclude our findings in Section~\ref{conclusion}.





\section{Sample Construction} \label{sample_construction}

\subsection{Field M Dwarf Sample}

The field M dwarf sample is constructed based on the SpeX spectroscopic survey carried out by \cite{Terrien2012} between 2011 and 2013. The whole stellar sample consisted of 886 nearby M dwarfs mostly selected according to their proper motions \citep{Lepine2005,Lepine2011} with a small set of planet hosts as well as wide binaries that serve as abundance
calibration stars \citep{RojasAyala2010}. All targets were observed with SpeX in the short cross-dispersed (SXD) mode at a spectral resolving power of $\approx 2000$ covering the $JHK$ bands simultaneously. Later work from \cite{Terrien2015} provided the [Fe/H] and [M/H] measurements for most stars based on different calibrations (see Section~\ref{analysis}). We refer the readers to \cite{Terrien2015} for more details about the survey description, observations, and data reduction.

To obtain precise stellar properties, we crossmatch the full catalog with Gaia DR3 \citep{Gaia2023} through their 2MASS identifiers \citep{Cutri2003,skrutskie2006} to retrieve the stellar kinematic properties. We restrict the sample to stars with parallax and proper motion measurements from Gaia, which excluded 109 stars. We do not make a specific cut on the significance of these measurements as most stars are nearby and have well-determined parallax and proper motions. We find that 24 stars in \cite{Terrien2015} do not have metallicity measurements, thus we remove them from our sample. To minimize the effect of giant planets in the field M dwarf sample, we filter out and exclude 7 known giant planet M dwarf hosts, all of which are included in our planet sample below, thus the final field M star sample ends with 746 stars. The full catalog is available in Table~\ref{field_M_star_sample}. We note that we do not have prior information on whether the rest of the stars in the field M dwarf sample host giant planets since most of them do not have radial velocity follow-up observations. Therefore, the field M dwarf group could be regarded as a mixed sample, containing stars without gas giants as well as a small fraction of stars hosting giant planets. Here, we demonstrate that such a contamination will not affect the final conclusion. According to the occurrence rate measurements, we conservatively estimate that there are about two M dwarfs hosting hot Jupiters and twenty M stars harboring cold Jupiters in the field M dwarf sample \citep{Sabotta2021,Gan2023,Bryant2023}. We randomly choose 22 stars in the field M dwarf sample, assuming they host giant planets and have the highest metallicity (0.7 dex) as in our planet sample. We repeat the simulation above for 5000 times and we find the metallicity distribution almost does not change with the median [Fe/H] varying between -0.07 and -0.06 dex for the $H$-band measurements, 0.07 and 0.08 dex for the $K$-band measurements as well as [M/H] ranging between -0.04 and -0.02 dex for the $H$-band results, 0.0 and 0.02 dex for the $K$-band results, all of which are within $1\sigma$.



\subsection{M Dwarf Giant Planet Sample}

\begin{figure*}
\centering
\includegraphics[width=0.99\textwidth]{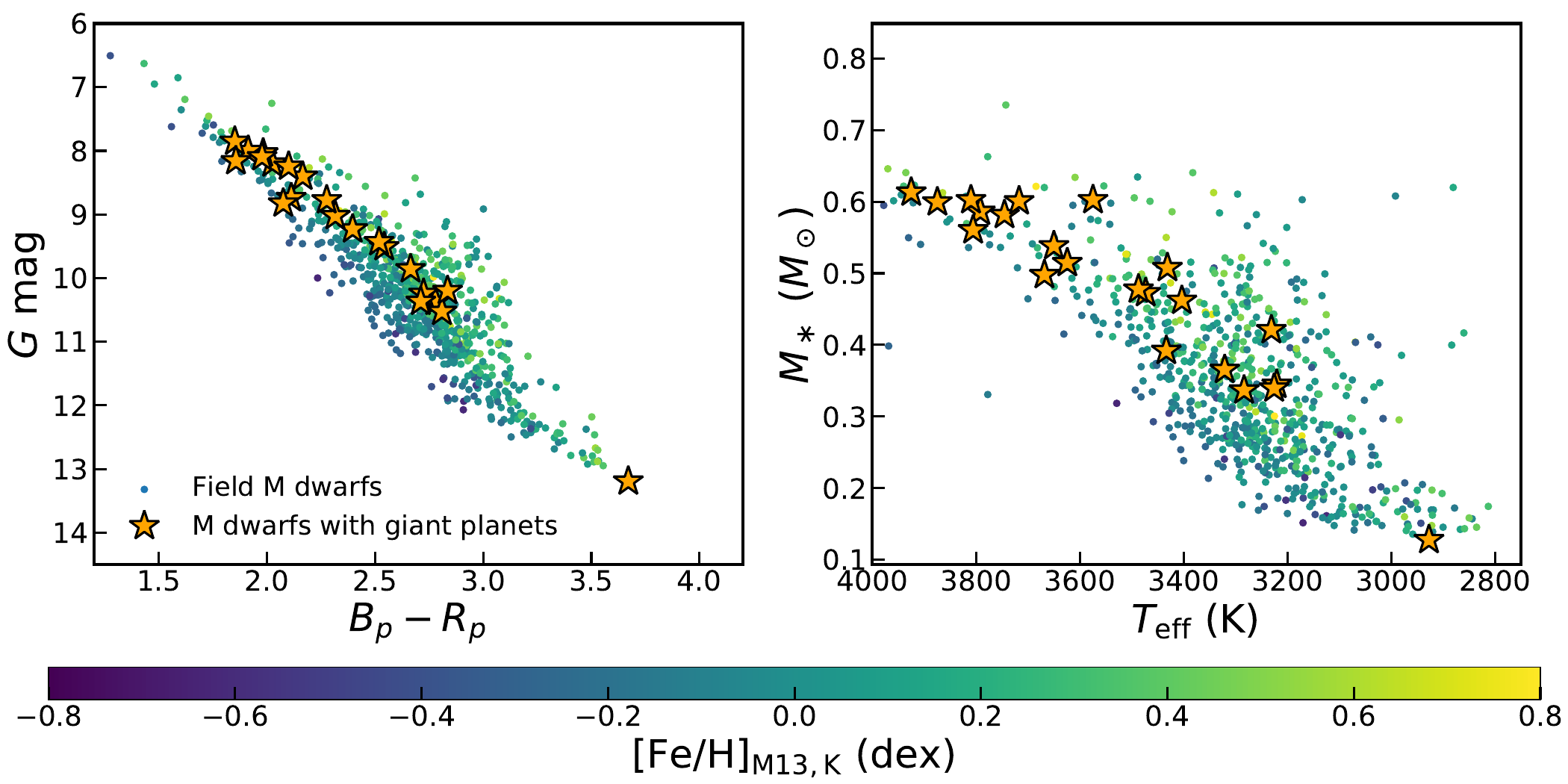}
\caption{\textit{Left panel:} The Gaia color–magnitude diagram of 746 field M dwarfs colored by their [Fe/H], and 22 M stars with confirmed giant planets marked as orange stars. \textit{Right panel:} The refined stellar masses and effective temperatures of two samples (see Section~\ref{analysis} for details).}
\label{stellar_parameters}
\end{figure*}

We constructed the M dwarf giant planet sample with the help of information from the NASA Exoplanet Archive \citep{Akeson2013} accessed on 2023 April 1st. We accepted all M dwarfs with effective temperature $T_{\rm eff} < 4000$ K and stellar mass $M_{\ast}< 0.65\ M_\odot$ that have confirmed giant planets with mass $0.2\ M_{\rm Jup} \leq M_p<13.6\ M_\mathrm{Jup}$ detected by either transit or radial velocity surveys. The raw selection during this step is based on stellar parameters from the literature, thus from different sources (i.e., spectroscopic facilities and methodologies). Both stellar and planet parameters are refined in a homogeneous way (see Section~\ref{analysis}). After excluding objects that are beyond the sky coverage of SpeX (Dec$<-30^{\circ}$), a total of 22 M dwarfs including 5 multi-giant-planet systems (GJ 1148, GJ 317, GJ 3512, GJ 849 and GJ 876) are left. Figure~\ref{stellar_parameters} shows the color-magnitude diagram and the refined stellar mass versus effective temperature distribution of the two samples. Our planet sample eventually contains these 22 M stars with 27 giant planets. 

We emphasize that our planet sample includes planetary systems detected by different surveys, and the potential bias between different selection functions is challenging to characterize. Therefore, we are not able to probe the giant planet occurrence rates in different stellar metallicity bins as other works \citep[e.g.,][]{Fischer2005}. Instead, we treat the M dwarf planet sample and the field M dwarf group as independent populations and test whether their metallicities could be drawn from a single distribution. The reason is that M dwarfs with confirmed giant planets are rare. For a specific transit or RV survey, the total number of detected systems is limited, prohibiting attempts to compare the occurrence rates in several metallicity bins due to large Poisson noise. 

\section{SpeX Observations}\label{observation}

\begin{figure*}
\centering
\includegraphics[width=0.75\textwidth]{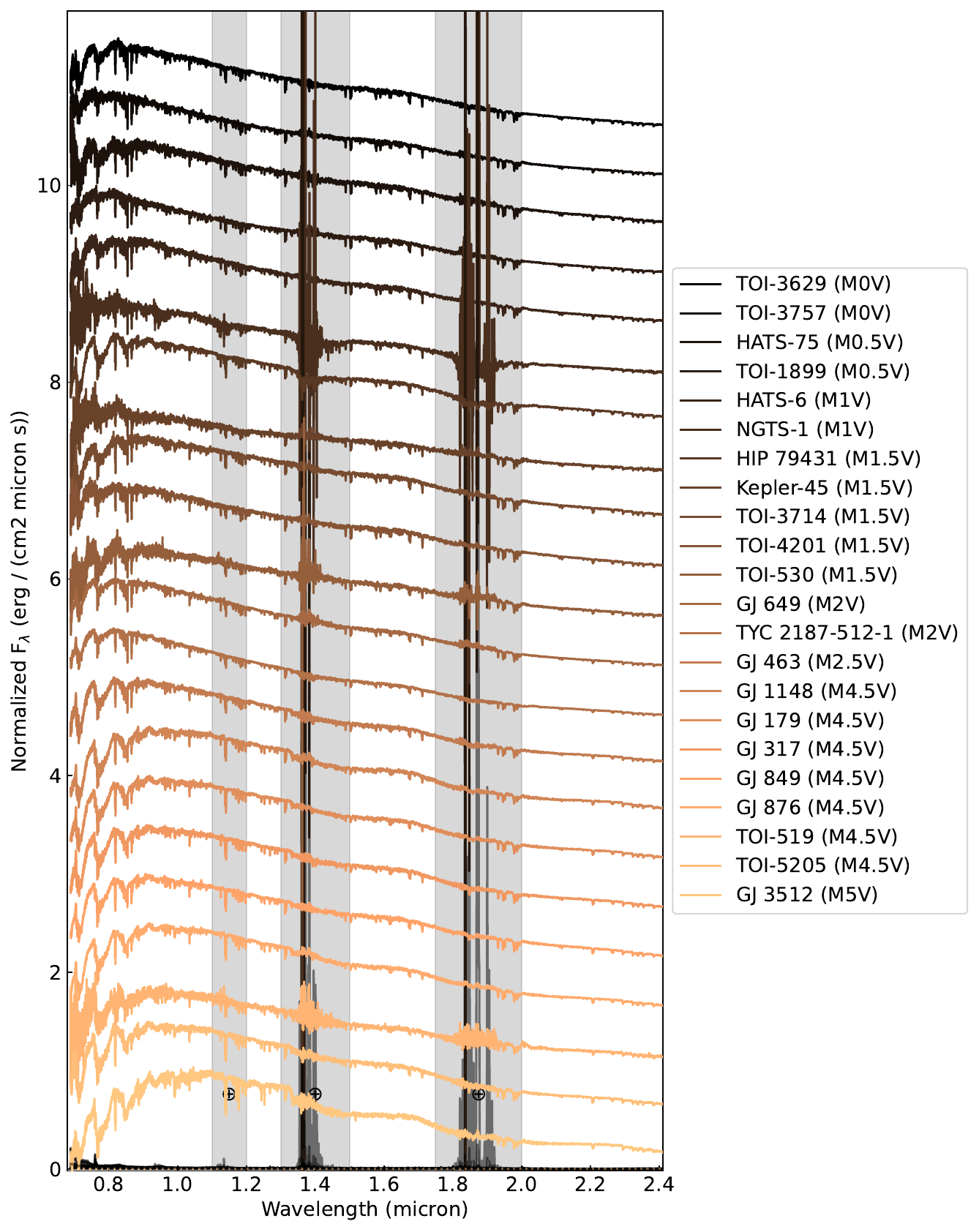}
\caption{The normalized SpeX spectra of 22 M dwarfs with confirmed giant planets. The data are presented in the order of stellar types derived through spectrum match (see Section~\ref{analysis}). Three vertical grey areas mark telluric absorption features, which we excluded during the analysis. All SpeX spectra we collected are available as the Data behind the Figure.}
\label{spectra}
\end{figure*}

We collected at least two ABBA nod sequences (8 exposures) for every giant planet host star\footnote{The SpeX spectra of three targets NGTS-1, TOI-519 and TOI-530 were retrieved from \cite{Gore2024}, which were taken using the same mode as here and \cite{Terrien2012}.} in our planet sample on 2023 August 9th and 11th, and December 28th using SpeX with the same observation configuration as \cite{Terrien2012}, under the program 2023B078. We applied the short cross-dispersed (SXD) mode with the $0\farcs3 \times 15\arcsec$ slit aligned with the parallactic angle at a spectral resolving power of $\approx 2000$, spanning a wavelength range of 0.8--2.4~$\mu$m. After the target observation, we obtained the spectrum of a nearby standard A0 star at an equivalent airmass for flux and telluric calibrations, followed by arc lamp and flat field lamp exposures. All spectroscopic data were then reduced using \code{SpeXtool} v4.1 \citep{Cushing2004} using standard settings and shifted to the rest frame. The resulting merged spectra have a signal-to-noise ratio (SNR) $\gtrsim 100$, all of which are shown in Figure~\ref{spectra}. 

\section{Analysis}\label{analysis}

\subsection{Metallicity}\label{metallicity}

\begin{figure}
\centering
\includegraphics[width=0.49\textwidth]{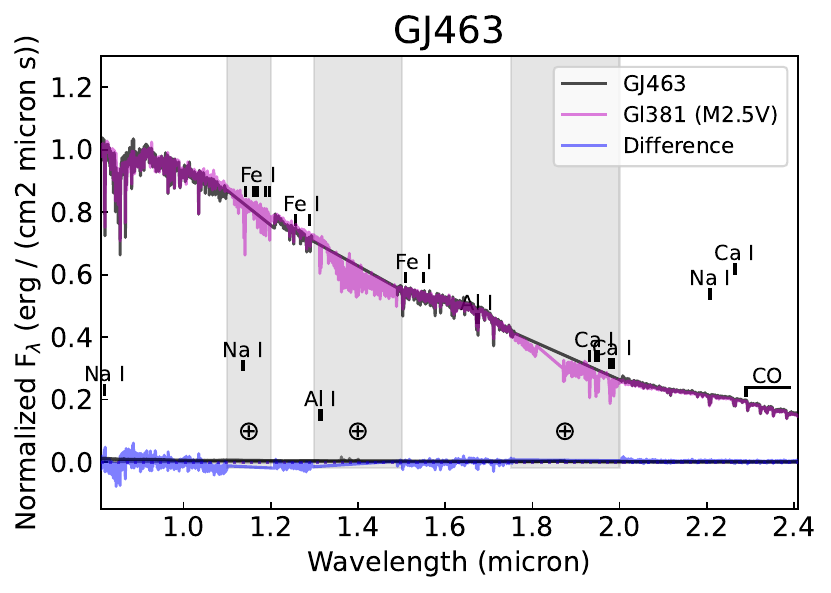}
\caption{An example plot of spectrum match in Section~\ref{analysis}. The black solid line is the normalized SpeX near-infrared spectrum of the target star GJ 463 while the best-match comparison spectrum taken from the IRTF library \citep{Rayner2009} is shown in magenta. Strong atomic and molecular features are marked. The residuals (blue) are presented below. }
\label{classify}
\end{figure}

We first determined the spectral types for the planet sample by comparing the 1d merged spectra to the IRTF spectral library \citep{Cushing2005,Rayner2009}. We utilized the \code{SPLAT} code \citep{Burgasser2017} to normalize the data and find the best matches through $\chi^{2}$ minimization. In the match, we exclude the regions close to 1.1--1.2, 1.3--1.5 and 1.75--2.0~$\mu m$ with strong telluric absorption that may dominate the residuals. We show an example of our spectrum match in Figure~\ref{classify}. 

The spectral types of the field M dwarf sample come from \cite{Terrien2015}, which instead applied the NIR calibration developed in \cite{Newton2014} based on the $\rm H_{2}O$-$\rm K2$ index defined by \cite{RojasAyala2012}. We examined the consistency of spectral type (SpT) between two methods with the 7 overlapped giant planet host stars and we find that $\Delta_{\rm SpT}\leq 1.0$. We thus neglected the difference and used them as inputs from the same source in the following steps. The spectral type of our planet sample ranges from M0V to M5V while the field M dwarf sample spans M0V to M5.5V. 

Building on the derived spectral types, we measured the iron abundance [Fe/H] and overall metallicity [M/H] following the same method as in \cite{Terrien2015}. We made use of the \code{metal} algorithm \citep{Mann2013} to determine the metallicity based on the $H$-band \citep{Terrien2012} and $K$-band \citep{RojasAyala2012} relations. The methodology is calibrated with a sample of wide binaries containing a Sun-like primary and a low-mass M star secondary, by determining the correlation between the metallicity of the central star and the features in the companion dwarf spectra \citep{Mann2013}. Since the $J$-band calibration has a larger scatter, we chose not to use it in this work. We note that the $K$-band spectra are expected to provide more accurate metallicity results because they are less affected by the telluric contamination than $H$-band spectra \citep{Dressing2019}. The systematic errors of $H$- and $K$-band calibrations are 0.09 and 0.08 dex, which are considered in the following analysis. 

\begin{figure*}
\centering
\includegraphics[width=0.99\textwidth]{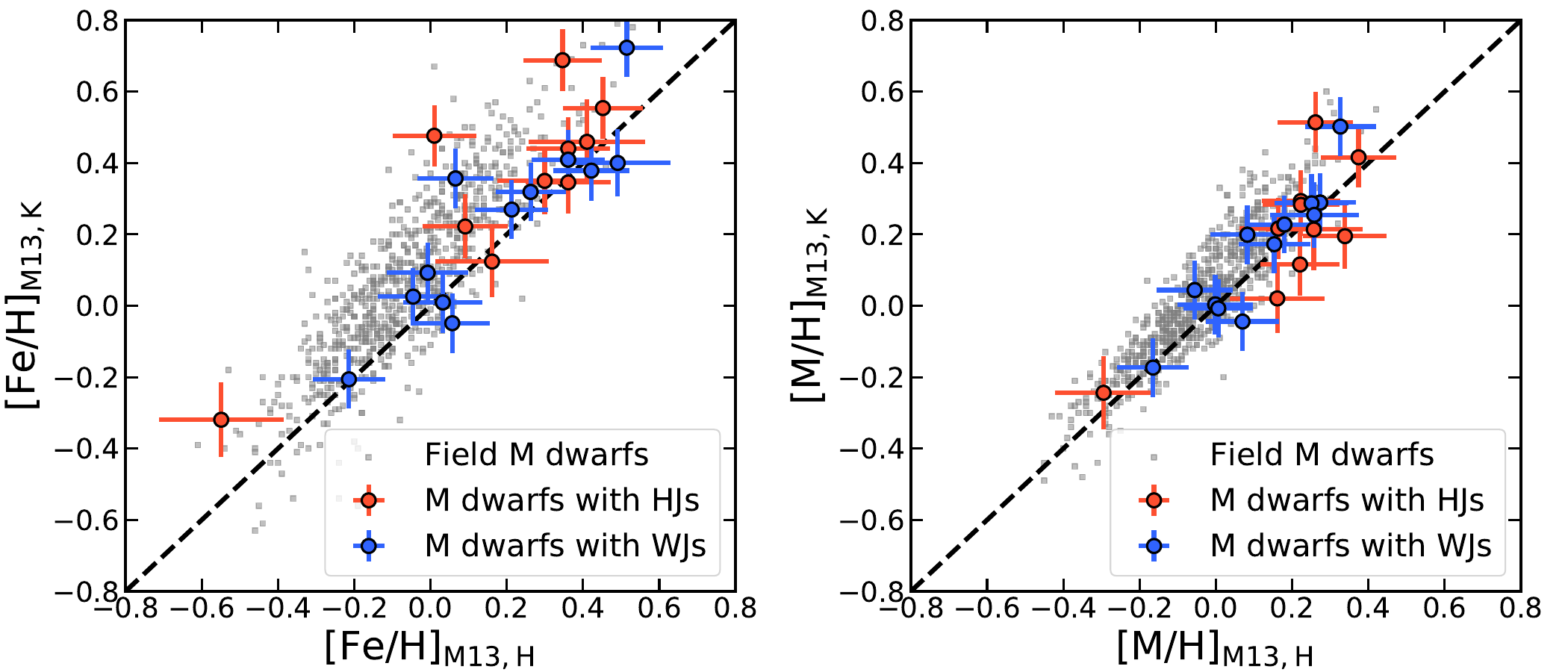}
\caption{Comparisons of [Fe/H] (left) and [M/H] (right) between $H$-band and $K$-band measurements. The background grey dots are the field M stars. The red and blue dots are the M dwarfs with hot and warm Jupiters with scaled semi-major axis $a/R_\ast\leq 20$ and $a/R_\ast > 20$, respectively. The systematic errors of 0.09 and 0.08 dex of the $H$ and $K$ band calibration have been taken into account. The black dashed line marks the one-to-one function.}
\label{HK_comparison}
\end{figure*}

Figure~\ref{HK_comparison} presents the comparisons of [Fe/H] and [M/H] outputs from $H$ and $K$ band spectra with subscripts of \textit{M13,H} and \textit{M13,K}. Overall, we find that the results agree well with each other although the $K$-band outputs appear to be slightly higher than those from $H$-band toward the high metallicity end. The discrepancy might be due to the wavelength calibration or the data reduction procedure but the exact reason for the offset is still unclear. Therefore, we chose to independently investigate the metallicity distribution based on the results from two bands. We caution the readers these calibrations were untested for M dwarfs beyond M5 and metallicity greater than 0.56 dex and the corresponding outputs could be interpreted as extrapolations since the wide binary sample used in \cite{Mann2013} did not have stars outside this parameter space. The majority of stars in both our planet sample and field M dwarf sample are within these ranges (see Figure~\ref{HK_comparison}) so we consider the outliers are not expected to significantly impact the final results. 

\subsection{Refined Stellar and Planet Properties}

In this section, we obtain the stellar properties of our planet sample in a homogeneous way to refine the planet parameters, which enables the investigation of the relation between stellar metallicity and other planet properties. In short, we apply the empirical relations, calibrated with nearby binaries, between stellar physical parameters and absolute magnitude as well as color. 

Combining the parallax from Gaia DR3 \citep{Gaia2023} and $m_{K}$ magnitude from 2MASS \citep{Cutri2003,skrutskie2006}, we first computed the absolute magnitude $M_K$ of each M dwarf in our sample. The stellar masses $M_{\ast}$ were then estimated through the $M_{\ast}$-$M_{K}$ relation\footnote{\url{https://github.com/awmann/M_-M_K-}} derived by \cite{Mann2019}, which is suitable for stars spanning a mass range of $0.075 \leq M_\ast \leq 0.7\ M_\odot$ with a systematic uncertainty of about 3\%. Since the effect of [Fe/H] on the $M_{\ast}$-$M_{K}$ relation is weak, about $0.0\pm2.2$\% change in mass per dex change in [Fe/H] for stars in the solar neighborhood \citep{Mann2019}, we do not take metallicity into account here. In terms of stellar radius $R_\ast$, we employed the third-order polynomial relation between $R_{\ast}$ and $M_K$ \citep[see Eq. 4 in][]{Mann2015}. This relation is valid for stars that have spectral types ranging from K7 to M7 with a precision of 3\%. 

Furthermore, we estimated the stellar effective temperature using the $T_{\rm eff}$-color relation reported in \cite{Mann2015}. We retrieved the $V$-band magnitudes from APASS \citep{Henden2016} and $JH$-magnitudes from 2MASS \citep{Cutri2003,skrutskie2006}. We opted to use the relation that includes two stellar color terms $V-J$ and $J-H$, which gives a scatter of about 50~K. Combined with the typical spectroscopic
error, we conservatively adopt a systematic uncertainty of 70~K. The final uncertainties of $M_\ast$, $R_\ast$, and $T_{\rm eff}$ are determined through error propagation after accounting for the aforementioned systematic errors of the methodology. We provide the information of the field M star catalog in Table~\ref{field_M_star_sample} with each column described in Table~\ref{table_column}. The stellar properties of our planet sample are listed in Table~\ref{giant_planet_M_star_sample}. 

Finally, we recalculated the planetary mass $M_p$, semi-major axis $a$, and equilibrium temperature $T_{\rm eq}$ using the updated stellar properties above together with the orbital results from the literature. For each planet, we randomly draw a set of parameters including effective temperature $T_{\rm eff}$, stellar mass $M_\ast$ and radius $R_\ast$, period $P$, eccentricity $e$, and radial velocity semi-amplitude $K$ from normal distributions, centered at the best-fits with $\sigma$ adopted as the higher value of their lower and upper uncertainty to be conservative. We fixed the eccentricity at zero if the published result is 1) smaller than 0.1 and consistent with 0 within $3\sigma$; or 2) only an upper limit. We repeated the sampling process 5000 times, and recorded the median and standard deviation of each distribution as the final value and uncertainty of planetary parameters. Here, we do not induce the constraint on the scaled semi-major axis $a/R_\ast$ from the light curve so the uncertainty is higher for transiting systems compared with the value in literature. Table~\ref{giant_planet_properties} summarizes the refined planet properties of our planet sample.

\begin{sidewaystable*}\tiny
    \centering
    \vspace{10em}
    {\renewcommand{\arraystretch}{1.05}
    \caption{Field M Dwarf Catalog. The descriptions of each column are presented in Table~\ref{table_column}.}
    \begin{tabular}{lccccccccccccccc}
        \hline\hline
        ID   &$\mu_\alpha$ &$\mu_\delta$ &$\varpi$ &$V$ &$J$ &$H$ &$K$ &SpT &MFeHH  &MFeHK  &MMHH  &MMHK &$M_\ast$ &$R_{\ast}$ &$T_{\rm eff}$ \\\hline
J00063925-0705354	&$-104.95\pm0.43$	&$119.32\pm0.26$	&$46.96\pm0.40$	&$14.71\pm0.03$	&$9.83\pm0.03$	&$9.26\pm0.03$	&$8.96\pm0.02$	&5.46	&-0.31	&-0.02	&-0.17	&-0.08	&$0.25\pm0.02$	&$0.28\pm0.02$	&$3078\pm83$	\\
J00102561+6212374	&$-28.68\pm0.02$	&$31.44\pm0.02$	&$37.88\pm0.02$	&$13.80\pm0.02$	&$9.65\pm0.02$	&$9.09\pm0.02$	&$8.81\pm0.01$	&3.49	&-0.11	&0.17	&-0.04	&0.06	&$0.33\pm0.02$	&$0.35\pm0.02$	&$3288\pm79$	\\
J00115302+2259047	&$119.50\pm0.07$	&$-212.39\pm0.07$	&$48.83\pm0.06$	&$13.00\pm0.04$	&$8.86\pm0.02$	&$8.31\pm0.04$	&$7.99\pm0.02$	&3.62	&0.09	&0.25	&0.04	&0.14	&$0.37\pm0.02$	&$0.38\pm0.02$	&$3285\pm83$	\\
J00161607+1951515	&$708.14\pm0.04$	&$-748.88\pm0.04$	&$65.05\pm0.04$	&$13.55\pm0.20$	&$8.89\pm0.03$	&$8.34\pm0.04$	&$8.10\pm0.03$	&4.44	&-0.03	&-0.07	&-0.15	&-0.09	&$0.27\pm0.02$	&$0.29\pm0.02$	&$3130\pm102$	\\
J00165629+0507261	&$-89.93\pm0.04$	&$-623.93\pm0.03$	&$56.16\pm0.04$	&$13.81\pm0.02$	&$9.40\pm0.02$	&$8.87\pm0.05$	&$8.59\pm0.02$	&4.52	&-0.17	&-0.1	&-0.19	&-0.11	&$0.25\pm0.02$	&$0.27\pm0.02$	&$3190\pm85$	\\
J00182256+4401222	&$2891.52\pm0.01$	&$411.83\pm0.01$	&$280.71\pm0.02$	&$8.09\pm0.10$	&$5.25\pm0.26$	&$4.48\pm0.20$	&$4.02\pm0.02$	&1.89	&-0.23	&-0.28	&-0.25	&-0.18	&$0.40\pm0.02$	&$0.41\pm0.03$	&$3968\pm307$	\\
J00182549+4401376	&$2862.80\pm0.02$	&$336.43\pm0.02$	&$280.69\pm0.03$	&$11.33\pm0.20$	&$6.79\pm0.02$	&$6.19\pm0.02$	&$5.95\pm0.02$	&4.10	&-0.13	&-0.13	&-0.14	&-0.1	&$0.16\pm0.02$	&$0.20\pm0.02$	&$3200\pm99$	\\
J00243478+3002295	&$585.89\pm0.05$	&$9.96\pm0.03$	&$51.32\pm0.04$	&$14.52\pm0.07$	&$9.78\pm0.02$	&$9.22\pm0.02$	&$8.89\pm0.02$	&4.70	&0.17	&0.23	&0.16	&0.1	&$0.23\pm0.02$	&$0.26\pm0.02$	&$3115\pm82$	\\
J00252063+2253121	&$-234.41\pm0.04$	&$-463.95\pm0.03$	&$61.90\pm0.03$	&$14.25\pm0.06$	&$9.72\pm0.02$	&$9.15\pm0.02$	&$8.87\pm0.02$	&4.71	&-0.11	&-0.01	&-0.18	&-0.04	&$0.19\pm0.02$	&$0.23\pm0.02$	&$3182\pm81$	\\
J00270673+4941531	&$367.63\pm0.02$	&$-224.69\pm0.02$	&$44.81\pm0.03$	&$14.19\pm0.04$	&$9.73\pm0.02$	&$9.16\pm0.02$	&$8.85\pm0.02$	&4.00	&0.14	&0.25	&0.07	&0.15	&$0.27\pm0.02$	&$0.30\pm0.02$	&$3202\pm80$	\\
J00283948-0639481	&$-327.79\pm0.05$	&$-800.63\pm0.03$	&$74.71\pm0.04$	&$12.50\pm0.20$	&$8.04\pm0.02$	&$7.50\pm0.06$	&$7.19\pm0.02$	&4.18	&-0.25	&0.02	&-0.13	&-0.02	&$0.35\pm0.02$	&$0.37\pm0.02$	&$3182\pm106$	\\
J00294322+0112384	&$-171.70\pm0.03$	&$-147.50\pm0.03$	&$32.25\pm0.03$	&$12.28\pm0.02$	&$9.15\pm0.02$	&$8.54\pm0.04$	&$8.31\pm0.02$	&1.75	&-0.27	&-0.16	&-0.24	&-0.15	&$0.46\pm0.03$	&$0.47\pm0.03$	&$3700\pm84$	\\
J00312156+0009294	&$522.22\pm0.03$	&$111.05\pm0.02$	&$32.90\pm0.03$	&$13.62\pm0.02$	&$9.76\pm0.02$	&$9.16\pm0.03$	&$8.90\pm0.02$	&3.23	&-0.09	&0.12	&-0.05	&0.03	&$0.36\pm0.02$	&$0.38\pm0.02$	&$3408\pm81$	\\
J00313539-0552115	&$350.97\pm0.04$	&$-1055.47\pm0.03$	&$71.02\pm0.03$	&$12.87\pm0.20$	&$8.76\pm0.03$	&$8.22\pm0.03$	&$7.95\pm0.02$	&3.68	&-0.24	&-0.11	&-0.22	&-0.1	&$0.26\pm0.02$	&$0.29\pm0.02$	&$3288\pm103$	\\
J00321574+5429027	&$-208.27\pm0.02$	&$-441.04\pm0.02$	&$49.95\pm0.03$	&$13.86\pm0.03$	&$9.39\pm0.02$	&$8.83\pm0.02$	&$8.57\pm0.01$	&4.49	&0.0	&-0.04	&-0.08	&-0.05	&$0.28\pm0.02$	&$0.30\pm0.02$	&$3193\pm80$	\\
J00325313-0434068	&$81.75\pm0.12$	&$-131.34\pm0.09$	&$52.85\pm0.10$	&$14.00\pm0.03$	&$9.28\pm0.02$	&$8.62\pm0.03$	&$8.35\pm0.02$	&4.83	&-0.04	&0.24	&0.1	&0.11	&$0.29\pm0.02$	&$0.31\pm0.02$	&$3189\pm81$	\\
J00383388+5127579	&$-228.57\pm0.03$	&$35.54\pm0.03$	&$55.83\pm0.03$	&$12.62\pm0.05$	&$8.89\pm0.03$	&$8.34\pm0.02$	&$8.06\pm0.02$	&2.83	&-0.28	&-0.17	&-0.23	&-0.15	&$0.32\pm0.02$	&$0.34\pm0.02$	&$3418\pm83$	\\
J00391896+5508132	&$-360.78\pm0.02$	&$-234.31\pm0.02$	&$58.19\pm0.02$	&$14.19\pm0.03$	&$10.02\pm0.02$	&$9.52\pm0.02$	&$9.24\pm0.02$	&4.05	&-0.32	&-0.4	&-0.38	&-0.32	&$0.17\pm0.02$	&$0.21\pm0.02$	&$3244\pm79$	\\
J00393349+1454188	&$331.91\pm0.03$	&$34.95\pm0.03$	&$34.82\pm0.03$	&$14.57\pm0.20$	&$9.96\pm0.03$	&$9.40\pm0.03$	&$9.12\pm0.02$	&4.01	&0.15	&0.15	&0.08	&0.09	&$0.31\pm0.02$	&$0.33\pm0.02$	&$3151\pm101$	\\
J00393374+1454348	&$327.54\pm0.25$	&$29.75\pm0.21$	&$34.26\pm0.22$	&$14.71\pm0.20$	&$9.83\pm0.03$	&$9.22\pm0.03$	&$8.95\pm0.02$	&4.67	&0.0	&0.28	&0.08	&0.15	&$0.34\pm0.02$	&$0.36\pm0.02$	&$3107\pm101$	\\
J00393379+6033153	&$-157.75\pm0.03$	&$-292.62\pm0.04$	&$27.12\pm0.03$	&$12.80\pm0.05$	&$9.20\pm0.02$	&$8.58\pm0.02$	&$8.34\pm0.02$	&2.42	&0.08	&0.31	&0.1	&0.19	&$0.52\pm0.03$	&$0.52\pm0.03$	&$3510\pm81$	\\
J00405623+3122565	&$-42.77\pm0.04$	&$-331.45\pm0.03$	&$44.39\pm0.03$	&$13.86\pm0.07$	&$9.49\pm0.02$	&$8.86\pm0.03$	&$8.59\pm0.02$	&3.33	&0.15	&0.56	&0.07	&0.38	&$0.31\pm0.02$	&$0.33\pm0.02$	&$3273\pm83$	\\
J00412078+5550044	&$324.33\pm0.01$	&$-73.26\pm0.02$	&$43.73\pm0.02$	&$14.77\pm0.20$	&$9.84\pm0.02$	&$9.31\pm0.02$	&$9.04\pm0.02$	&4.17	&-0.15	&-0.19	&-0.19	&-0.16	&$0.26\pm0.02$	&$0.28\pm0.02$	&$3038\pm98$	\\
J00442070+0907345	&$813.09\pm0.10$	&$-2.55\pm0.07$	&$40.01\pm0.08$	&$14.47\pm0.20$	&$9.50\pm0.03$	&$8.96\pm0.03$	&$8.62\pm0.02$	&4.48	&0.07	&0.19	&-0.01	&0.08	&$0.34\pm0.02$	&$0.36\pm0.02$	&$3033\pm100$	\\
J00462990+5038389	&$419.18\pm0.02$	&$-213.83\pm0.02$	&$30.95\pm0.02$	&$14.34\pm0.02$	&$9.96\pm0.02$	&$9.33\pm0.02$	&$9.07\pm0.02$	&3.25	&0.27	&0.44	&0.19	&0.32	&$0.36\pm0.02$	&$0.37\pm0.02$	&$3268\pm79$	\\
J00501752+0837341	&$66.96\pm0.07$	&$-35.13\pm0.04$	&$16.31\pm0.05$	&$14.35\pm0.04$	&$9.74\pm0.02$	&$9.15\pm0.02$	&$8.90\pm0.02$	&5.14	&-0.04	&-0.05	&-0.14	&-0.09	&$0.60\pm0.03$	&$0.62\pm0.03$	&$3171\pm80$	\\
J00523518+4712433	&$260.33\pm0.02$	&$76.84\pm0.02$	&$31.25\pm0.03$	&$13.98\pm0.20$	&$9.76\pm0.02$	&$9.19\pm0.02$	&$8.93\pm0.02$	&3.30	&-0.09	&-0.04	&-0.09	&-0.06	&$0.38\pm0.02$	&$0.39\pm0.02$	&$3275\pm100$	\\
J00563841+1727347	&$683.34\pm0.05$	&$-291.75\pm0.04$	&$54.85\pm0.04$	&$13.73\pm0.03$	&$9.28\pm0.02$	&$8.65\pm0.03$	&$8.37\pm0.03$	&3.78	&0.17	&0.57	&0.09	&0.37	&$0.28\pm0.02$	&$0.30\pm0.02$	&$3247\pm81$	\\
J00580115+3919111	&$-107.05\pm0.03$	&$25.00\pm0.02$	&$68.39\pm0.04$	&$14.17\pm0.03$	&$9.56\pm0.03$	&$8.95\pm0.03$	&$8.68\pm0.02$	&4.23	&-0.16	&-0.03	&-0.07	&-0.05	&$0.19\pm0.02$	&$0.22\pm0.02$	&$3186\pm83$	\\
J01004948+6656546	&$-236.24\pm0.02$	&$-54.67\pm0.02$	&$44.32\pm0.02$	&$13.43\pm0.02$	&$9.41\pm0.03$	&$8.73\pm0.03$	&$8.48\pm0.02$	&3.50	&0.12	&0.24	&0.1	&0.18	&$0.33\pm0.02$	&$0.35\pm0.02$	&$3415\pm83$	\\
$\cdots$ &$\cdots$ &$\cdots$ &$\cdots$ &$\cdots$ &$\cdots$ &$\cdots$ &$\cdots$ &$\cdots$ &$\cdots$ &$\cdots$ &$\cdots$ &$\cdots$ &$\cdots$ &$\cdots$ &$\cdots$\\
        \hline
    \label{field_M_star_sample}    
    \end{tabular}}
    \begin{tablenotes}
       \item[1]  (The full table including 746 stars is available in its entirety in the machine-readable format. A portion is shown here for guidance regarding its form and content.) 
    \end{tablenotes}
\end{sidewaystable*}

\begin{table*}
    \centering
    {\renewcommand{\arraystretch}{1.03}
    \caption{Field M Dwarf Catalog Column Descriptions.}
    \begin{tabular}{llr}
        \hline\hline
        Name &Description &Ref. \\\hline
        ID &2MASS Identifier &2MASS$^{[1]}$\\
        $\mu_\alpha$ (mas yr$^{-1}$) & Gaia Proper Motion in R.A. &Gaia DR3$^{[2]}$\\
        $\mu_\delta$ (mas yr$^{-1}$) & Gaia Proper Motion in Decl. &Gaia DR3\\
        $\varpi$ (mas) &Gaia Parallax &Gaia DR3\\
        $V$ (mag) &APASS $V$ Band Magnitude &APASS$^{[3]}$\\
        $J$ (mag) &2MASS $J$ Band Magnitude &2MASS\\
        $H$ (mag) &2MASS $H$ Band Magnitude &2MASS\\
        $K$ (mag) &2MASS $K$ Band Magnitude &2MASS\\
        SpT &M Spectral Subtype defined in \cite{Newton2014} &\cite{Terrien2015}\\
        MFeHH (dex) &[Fe/H] from H-band defined in \cite{Mann2013}, denoted as $\rm [Fe/H]_{\rm M13,H}$ &\cite{Terrien2015}\\
        MFeHK (dex) &[Fe/H] from K-band defined in \cite{Mann2013}, denoted as $\rm [Fe/H]_{\rm M13,K}$&\cite{Terrien2015}\\
        MMHH (dex) &[M/H] from H-band defined in \cite{Mann2013}, denoted as $\rm [M/H]_{\rm M13,H}$ &\cite{Terrien2015}\\
        MMHK (dex) &[M/H] from K-band defined in \cite{Mann2013}, denoted as $\rm [M/H]_{\rm M13,K}$ &\cite{Terrien2015}\\
        $M_\ast$ ($M_\odot$) &Stellar Mass from $M_{\ast}$-$M_{K}$ Relation in \cite{Mann2019} &This Work\\
        $R_\ast$ ($R_\odot$) &Stellar Radius from $R_{\ast}$-$M_{K}$ Relation in \cite{Mann2015} &This Work\\
        $T_{\rm eff}$ ($K$) &Stellar Effective Temperature from $T_{\rm eff}$-Color Relation in \cite{Mann2015}$^{[4]}$ &This Work\\
        \hline
    \label{table_column}    
    \end{tabular}}
    \begin{tablenotes}
       \item[1]  [1] \cite{Cutri2003,skrutskie2006}; [2] \cite{Gaia2023}; [3] \cite{Henden2016}; [4] We use the empirical relation that includes two color terms: $V-J$ and $J-H$. 
    \end{tablenotes}
\end{table*}

\begin{table*}\scriptsize
    \centering
    {\renewcommand{\arraystretch}{1.03}
    \caption{Stellar Properties of the M Dwarf Giant Planet Sample.}
    \begin{tabular}{llcccccccc}
        \hline\hline
        Target (2MASS ID)  &Other Identifier     &SpT &$\rm [Fe/H]_{M13,H}$  &$\rm [Fe/H]_{M13,K}$  &$\rm [M/H]_{M13,H}$  &$\rm [M/H]_{M13,K}$ &$M_\ast\ (M_\odot)$ &$R_{\ast}\ (R_\odot)$ &$T_{\rm eff}\ (K)$\\\hline
J11414471+4245072	&GJ 1148$^{[1]}$	&4.5	&$0.03\pm0.10$	&$0.01\pm0.08$	&$-0.00\pm0.10$	&$0.00\pm0.08$	&$0.34\pm0.02$	&$0.36\pm0.02$	&$3220\pm80$	\\
J04520573+0628356	&GJ 179$^{[2]}$	&4.5	&$0.21\pm0.09$	&$0.27\pm0.08$	&$0.15\pm0.09$	&$0.17\pm0.08$	&$0.36\pm0.02$	&$0.38\pm0.02$	&$3321\pm86$	\\
J08405923-2327232	&GJ 317$^{[3]}$	&4.5	&$0.07\pm0.10$	&$0.36\pm0.08$	&$0.08\pm0.09$	&$0.20\pm0.08$	&$0.42\pm0.03$	&$0.43\pm0.03$	&$3231\pm110$	\\
J08412013+5929505	&GJ 3512$^{[4]}$	&5.0	&$-0.01\pm0.11$	&$0.09\pm0.08$	&$-0.06\pm0.10$	&$0.04\pm0.08$	&$0.13\pm0.02$	&$0.16\pm0.02$	&$2927\pm94$	\\
J12230024+6401506	&GJ 463$^{[5]}$	&2.5	&$0.06\pm0.10$	&$-0.05\pm0.08$	&$0.07\pm0.09$	&$-0.04\pm0.08$	&$0.47\pm0.03$	&$0.47\pm0.03$	&$3473\pm83$	\\
J16580884+2544392	&GJ 649$^{[6]}$	&2.0	&$-0.05\pm0.09$	&$0.03\pm0.08$	&$0.01\pm0.09$	&$-0.01\pm0.08$	&$0.51\pm0.03$	&$0.52\pm0.03$	&$3624\pm91$	\\
J22094029-0438267	&GJ 849$^{[7]}$	&4.5	&$0.36\pm0.09$	&$0.41\pm0.08$	&$0.27\pm0.09$	&$0.29\pm0.08$	&$0.46\pm0.03$	&$0.46\pm0.03$	&$3403\pm90$	\\
J22531672-1415489	&GJ 876$^{[8]}$	&4.5	&$0.26\pm0.09$	&$0.32\pm0.08$	&$0.18\pm0.09$	&$0.23\pm0.08$	&$0.34\pm0.02$	&$0.35\pm0.02$	&$3283\pm90$	\\
J05523523-1901539	&HATS-6$^{[9]}$	&1.0	&$0.36\pm0.10$	&$0.35\pm0.09$	&$0.16\pm0.10$	&$0.22\pm0.08$	&$0.59\pm0.03$	&$0.60\pm0.03$	&$3791\pm94$	\\
J04034783-2524320	&HATS-75$^{[10]}$	&0.5	&$0.30\pm0.12$	&$0.35\pm0.09$	&$0.34\pm0.10$	&$0.19\pm0.09$	&$0.58\pm0.03$	&$0.59\pm0.03$	&$3745\pm110$	\\
J16124178-1852317	&HIP 79431$^{[11]}$	&1.5	&$0.51\pm0.09$	&$0.72\pm0.08$	&$0.33\pm0.09$	&$0.50\pm0.08$	&$0.48\pm0.03$	&$0.48\pm0.03$	&$3487\pm87$	\\
J19312949+4103513	&Kepler-45$^{[12]}$	&1.5	&$0.41\pm0.15$	&$0.46\pm0.12$	&$0.26\pm0.13$	&$0.21\pm0.11$	&$0.60\pm0.03$	&$0.62\pm0.03$	&$3810\pm110$	\\
J05305145-3637508	&NGTS-1$^{[13]}$	&1.0	&$-0.55\pm0.17$	&$-0.32\pm0.11$	&$-0.29\pm0.13$	&$-0.24\pm0.10$	&$0.56\pm0.03$	&$0.57\pm0.03$	&$3805\pm110$	\\
J19574239+4008357	&TOI-1899$^{[14]}$	&0.5	&$0.42\pm0.10$	&$0.38\pm0.08$	&$0.25\pm0.09$	&$0.29\pm0.08$	&$0.60\pm0.03$	&$0.62\pm0.03$	&$3574\pm110$	\\
J23591015+3918514	&TOI-3629$^{[15]}$	&0.0	&$0.45\pm0.10$	&$0.55\pm0.08$	&$0.37\pm0.10$	&$0.42\pm0.08$	&$0.60\pm0.03$	&$0.62\pm0.03$	&$3717\pm87$	\\
J04381253+3927299	&TOI-3714$^{[16]}$	&1.5	&$0.01\pm0.11$	&$0.48\pm0.08$	&$0.22\pm0.10$	&$0.29\pm0.08$	&$0.51\pm0.03$	&$0.51\pm0.03$	&$3431\pm101$	\\
J06040089+5501126	&TOI-3757$^{[17]}$	&0.0	&$0.09\pm0.11$	&$0.22\pm0.09$	&$0.22\pm0.10$	&$0.12\pm0.09$	&$0.62\pm0.03$	&$0.64\pm0.03$	&$3925\pm92$	\\
J06015391-1327410	&TOI-4201$^{[18]}$	&1.5	&$0.36\pm0.11$	&$0.44\pm0.09$	&$0.22\pm0.10$	&$0.28\pm0.08$	&$0.60\pm0.03$	&$0.62\pm0.03$	&$3874\pm92$	\\
J08182567-1939465	&TOI-519$^{[19]}$	&4.5	&$0.16\pm0.15$	&$0.12\pm0.10$	&$0.16\pm0.13$	&$0.02\pm0.09$	&$0.34\pm0.02$	&$0.36\pm0.02$	&$3225\pm100$	\\
J20550491+2421387	&TOI-5205$^{[20]}$	&4.5	&$0.35\pm0.10$	&$0.69\pm0.08$	&$0.26\pm0.10$	&$0.51\pm0.08$	&$0.39\pm0.02$	&$0.40\pm0.02$	&$3433\pm110$	\\
J06533906+1252545	&TOI-530$^{[21]}$	&1.5	&$0.49\pm0.14$	&$0.40\pm0.09$	&$0.26\pm0.11$	&$0.25\pm0.09$	&$0.54\pm0.03$	&$0.55\pm0.03$	&$3650\pm100$	\\
J21220626+2255531	&TYC 2187-512-1$^{[22]}$	&2.0	&$-0.21\pm0.09$	&$-0.21\pm0.08$	&$-0.16\pm0.09$	&$-0.17\pm0.08$	&$0.50\pm0.03$	&$0.50\pm0.03$	&$3668\pm86$	\\
        \hline
    \label{giant_planet_M_star_sample}    
    \end{tabular}}
    \begin{tablenotes}
       \item[1]  [1] \cite{Haghighipour2010,Trifonov2018,Rosenthal2021}; [2] \cite{Howard2010,Rosenthal2021}; [3] \cite{Johnson2007,Anglada2012GJ317,Rosenthal2021}; [4] \cite{Morales2019,Lopez2020}; [5] \cite{Endl2022,Sozzetti2023}; [6] \cite{Johnson2010,Rosenthal2021,Pinamonti2023}; [7] \cite{Butler2006,Bonfils2013,Rosenthal2021,Pinamonti2023}; [8] \cite{Marcy1998,Correia2010,Rivera2010,Trifonov2018,Rosenthal2021}; [9] \cite{Hartman2015}; [10] \cite{Jordan2022}; [11] \cite{Apps2010}; [12] \cite{Johnson2012,Bonomo2017}; [13] \cite{Bayliss2018}; [14] \cite{Canas2020,Lin2023}; [15] \cite{Canas2022,Hartman2023}; [16] \cite{Canas2022,Hartman2023}; [17] \cite{Kanodia2022}; [18] \cite{Gantoi4201,Hartman2023,Delamer2024}; [19] \cite{Parviainen2021,Kagetani2023,Hartman2023}; [20] \cite{Kanodia2023}; [21] \cite{Gantoi530}; [22] \cite{Quirrenbach2022}
    \end{tablenotes}
\end{table*}

\begin{table*}\scriptsize
    \centering
    {\renewcommand{\arraystretch}{1.05}
    \caption{Planet Properties of the M Dwarf Giant Planet Sample.}
    \begin{tabular}{lcccccc}
        \hline\hline
        Planet       &$M_p\sin i\ (M_{\rm Jup})$ &Mass Ratio  &$a\ ({\rm AU})$   &$a/R_\ast^{[1]}$ &$T_{\rm eq}\ (K)^{[2]}$ &Group$^{[3]}$\\\hline
        GJ 1148b &$0.297\pm0.007$ &$0.000823\pm0.000145$ &$0.1641\pm0.0017$ &$98.20\pm16.82$ &$229\pm20$ &WJ\\
GJ 1148c &$0.208\pm0.013$ &$0.000577\pm0.000107$ &$0.9010\pm0.0098$ &$539.20\pm92.40$ &$98\pm15$ &WJ\\
GJ 179b &$0.828\pm0.074$ &$0.002171\pm0.000416$ &$2.4266\pm0.0476$ &$1386.49\pm233.28$ &$63\pm12$ &WJ\\
GJ 317b &$1.753\pm0.037$ &$0.003973\pm0.000676$ &$1.1530\pm0.0111$ &$580.11\pm97.41$ &$94\pm15$ &WJ\\
GJ 317c &$1.644\pm0.044$ &$0.003726\pm0.000637$ &$5.2375\pm0.0888$ &$2635.17\pm444.01$ &$44\pm10$ &WJ\\
GJ 3512b &$0.474\pm0.011$ &$0.003534\pm0.000668$ &$0.3419\pm0.0040$ &$470.82\pm146.10$ &$95\pm15$ &WJ\\
GJ 3512c &$0.220\pm0.006$ &$0.001643\pm0.000311$ &$1.3491\pm0.0159$ &$1857.95\pm576.54$ &$47\pm10$ &WJ\\
GJ 463b &$1.500\pm0.139$ &$0.003027\pm0.000583$ &$3.4826\pm0.0732$ &$1575.16\pm236.52$ &$61\pm12$ &WJ\\
GJ 649b &$0.259\pm0.014$ &$0.000480\pm0.000086$ &$1.1155\pm0.0110$ &$463.54\pm61.33$ &$118\pm15$ &WJ\\
GJ 849b &$0.908\pm0.029$ &$0.001880\pm0.000327$ &$2.3415\pm0.0232$ &$1084.16\pm163.44$ &$73\pm13$ &WJ\\
GJ 849c &$1.002\pm0.047$ &$0.002075\pm0.000368$ &$4.9890\pm0.0745$ &$2309.97\pm349.20$ &$50\pm10$ &WJ\\
GJ 876b &$1.988\pm0.042$ &$0.005629\pm0.001001$ &$0.2115\pm0.0022$ &$129.07\pm25.31$ &$204\pm10$ &WJ\\
GJ 876c &$0.656\pm0.014$ &$0.001858\pm0.000330$ &$0.1322\pm0.0014$ &$80.69\pm15.82$ &$258\pm10$ &WJ\\
HATS-6b &$0.325\pm0.073$ &$0.000528\pm0.000150$ &$0.0365\pm0.0004$ &$13.09\pm2.08$ &$740\pm20$ &HJ\\
HATS-75b &$0.478\pm0.039$ &$0.000788\pm0.000153$ &$0.0323\pm0.0003$ &$11.79\pm1.89$ &$771\pm25$ &HJ\\
HIP 79431b &$2.068\pm0.058$ &$0.004138\pm0.000743$ &$0.3550\pm0.0040$ &$159.24\pm23.64$ &$195\pm15$ &WJ\\
Kepler-45b &$0.516\pm0.048$ &$0.000817\pm0.000161$ &$0.0301\pm0.0003$ &$10.48\pm2.18$ &$834\pm30$ &HJ\\
NGTS-1b &$0.770\pm0.053$ &$0.001305\pm0.000241$ &$0.0309\pm0.0003$ &$11.66\pm1.98$ &$787\pm25$ &HJ\\
TOI-1899b &$0.646\pm0.040$ &$0.001023\pm0.000188$ &$0.1564\pm0.0016$ &$54.29\pm7.09$ &$343\pm10$ &WJ\\
TOI-3629b &$0.249\pm0.023$ &$0.000395\pm0.000077$ &$0.0412\pm0.0004$ &$14.36\pm2.13$ &$693\pm18$ &HJ\\
TOI-3714b &$0.682\pm0.025$ &$0.001283\pm0.000226$ &$0.0260\pm0.0003$ &$10.97\pm1.68$ &$732\pm23$ &HJ\\
TOI-3757b &$0.263\pm0.028$ &$0.000405\pm0.000081$ &$0.0380\pm0.0004$ &$12.77\pm1.96$ &$776\pm21$ &HJ\\
TOI-4201b &$2.446\pm0.084$ &$0.003883\pm0.000685$ &$0.0387\pm0.0004$ &$13.50\pm2.21$ &$745\pm20$ &HJ\\
TOI-519b &$0.471\pm0.087$ &$0.001318\pm0.000335$ &$0.0160\pm0.0002$ &$9.65\pm2.17$ &$733\pm29$ &HJ\\
TOI-5205b &$1.068\pm0.048$ &$0.002614\pm0.000474$ &$0.0198\pm0.0002$ &$10.70\pm2.08$ &$740\pm28$ &HJ\\
TOI-530b &$0.405\pm0.086$ &$0.000713\pm0.000195$ &$0.0550\pm0.0006$ &$21.62\pm3.59$ &$554\pm17$ &WJ\\
TYC 2187-512-1b &$0.328\pm0.014$ &$0.000629\pm0.000113$ &$1.2137\pm0.0155$ &$521.48\pm85.95$ &$113\pm10$ &WJ\\
        \hline
    \label{giant_planet_properties}    
    \end{tabular}}
    \begin{tablenotes}
       \item[1]  [1]\ The uncertainties on the scaled semi-major axis $a/R_\ast$ of transiting systems are higher than literature values because here we do not use constraint from the light curve.  [2]\ We do not consider heat distribution between the dayside and nightside here and assume albedo $A_B=0$. [3]\ The hot Jupiter (HJ) and warm Jupiter (WJ) groups are designated as giant planets ($M_p>0.2\ M_\mathrm{Jup}$) with $a/R_\ast\leq 20$ and $a/R_\ast>20$.
    \end{tablenotes}
\end{table*}

\section{Discussions}\label{discussion}

\subsection{Giant Planets Favor Metal-Rich M Dwarfs}\label{Giant_Planets_Favor_Metal_Rich_M_Dwarfs}

With the [Fe/H] and [M/H] homogeneously measured based on spectral data from the same instrument through the same methodology, we investigate the metallicity distribution of the two samples we constructed. Since the behaviors of [Fe/H] and [M/H] are almost identical, we mainly focus on [Fe/H] in the following sections and put all results of [M/H] in the Appendix. 

Figure~\ref{HK_comparison_cumulative} illustrates the iron abundance distribution and the corresponding cumulative function of field M dwarfs (black line) and M stars with confirmed giant planets (orange line) from two bands. Each two-giant-planet system is treated as two single-planet systems independently. We also repeat the same analysis below but treat those M dwarfs hosting two giant planets as single stars, which gives similar results. By visual inspection, it appears that giant planets were mostly found around metal-rich M dwarfs, suggesting a correlation between their formation and stellar metallicity, similar to FGK counterparts \citep{Santos2004,Fischer2005,Johnson2010,Sousa2011,Maldonado2020,Osborn2020}. In addition, the iron abundance of M dwarfs harboring giant planets tends to show a bimodal distribution with one peak roughly located at the median [Fe/H] of field M dwarfs and the other at about 0.4 dex. However, the two-peak phenomenon is not seen in [M/H] (see Figure~\ref{MH_HK_comparison_cumulative}). Since the planet number in each [Fe/H] bin is small, we attribute the feature to Poisson noise and do not over-interpret it.

\begin{figure*}
\centering
\includegraphics[width=0.99\textwidth]{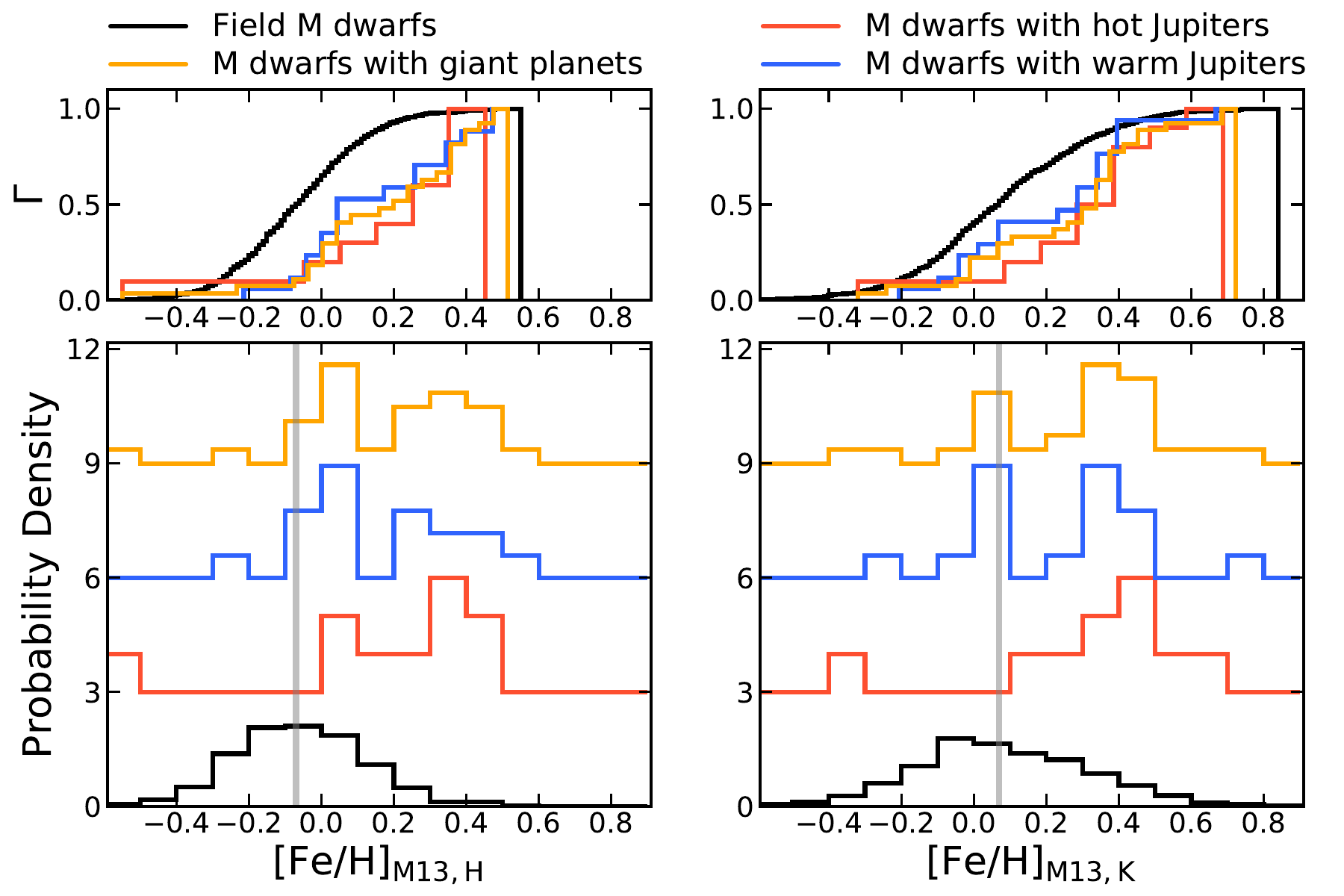}
\caption{Metallicity distribution of the field M stars (black) and M dwarfs with giant planets (orange), either hot Jupiters (red) with $a/R_\ast\leq 20$ or warm Jupiters (blue) with $a/R_\ast>20$. The left and right panels are the results from $H$- and $K$-band measurements. The vertical grey line in each panel marks the median [Fe/H] of the field M dwarf sample, $-$0.07 and 0.07 dex. The cumulative functions are shown above. Most confirmed giant planets were orbiting metal-rich M stars. The tentative bimodal distribution of [Fe/H] is likely due to Poisson noise.}
\label{HK_comparison_cumulative}
\end{figure*}

To statistically examine whether the field M dwarfs and M stars hosting giant planets have an identical underlying [Fe/H] distribution, we carry out Kolmogorov-Smirnov \citep[K-S;][]{Hodges1958} and Anderson-Darling \citep[A-D;][]{Scholz1987} tests. As the sample sizes are small, we apply the permutation method for the A-D test to obtain accurate $p$-values with the number of resamples set to $10^{6}$. We calculate the K-S and A-D statistic between two samples using the functions embedded in \code{SciPy} \citep{Virtanen2020}, which yield $p$-values of $4.9\times10^{-7}$ and $9.9\times 10^{-7}$ for the $H$-band as well as $6.3\times10^{-5}$ and $4.2\times 10^{-5}$ for the $K$-band metallicities, corresponding to $5\sigma$ and $4\sigma$ significance, respectively. To account for the uncertainty of each metallicity measurement, we resample the $\rm [Fe/H]_{M13,H}$ and $\rm [Fe/H]_{M13,K}$ by randomly drawing samples, assuming Gaussian distributions $\mathcal{N}{\rm ([Fe/H],\sigma_{[Fe/H]}^{2})}$. Since the metallicity uncertainty of the field star data set is unavailable in \cite{Terrien2015}, we assign 0.09 dex and 0.08 dex systematical errors\footnote{These systematical errors of the methodology dominate the final uncertainty according to our findings based on the planet sample (see Table~\ref{giant_planet_M_star_sample}).} to the $H$ and $K$ band measurements before randomizing. We loop the procedure 10000 times, record the $p$-values, and compute the fraction of trials with $p\leq 0.003$, corresponding to the $3\sigma$ significance criterion. We list all details of the K-S along with the A-D tests in Table~\ref{fehksadtest}. To sum up, we find that nearly all trials lead to $p$-values smaller than the 0.003 threshold, hence indicating a strong significance. Therefore, we conclude that the null hypothesis, the iron abundance distributions of field M dwarfs and M stars with giant planets are the same, could be rejected. Our findings suggest that giant planets favor metal-rich M stars, which agree with the conclusions from previous studies \citep{Johnson2009,RojasAyala2010}.

\subsection{Metallicity Preference of Hot and Warm Jupiters}\label{Metallicity_Preference_of_Hot_and_Warm_Jupiters}

Some initial efforts found that hot and warm Jupiters around M dwarfs may have different preferences of stellar metallicity \citep{Gantoi530,Gantoi4201}, hinting at different formation histories \citep{Pollack1996,Boss2002}. However, the large uncertainty of [Fe/H] as well as the systematic biases from heterogeneous instruments and methodologies that are challenging to characterize impede these works to draw a robust conclusion. Here, we revisit the puzzle using our samples characterized through a uniform pathway.

We divide the full giant planet sample into two subclasses: hot and warm Jupiters based on the refined planet properties. Following \cite{Gantoi4201}, we designate hot Jupiters (HJ) as planets having $M_{p}\geq 0.2\ M_\mathrm{Jup}$ and $a/R_\ast \leq 20$ and warm Jupiters (WJ) as planets within the same mass range but $a/R_\ast > 20$. In this way, we find 10 HJs and 17 WJs, including 5 multi-warm Jupiter systems. Among two groups, there are one HJ and one WJ whose host stars plausibly have sub-solar metallicities: NGTS-1 ($\rm [Fe/H]_{M13,H}=-0.55\pm0.17$~dex, $\rm [Fe/H]_{M13,K}=-0.32\pm0.11$~dex)\footnote{For NGTS-1, \cite{Bayliss2018} fixed an overall metallicity of $\rm [M/H] = 0$ dex in the analysis but did not have an estimate on $\rm [Fe/H]$.} and TYC 2187-512-1 ($\rm [Fe/H]_{M13,H}=-0.21\pm0.09$~dex, $\rm [Fe/H]_{M13,K}=-0.21\pm0.08$~dex)\footnote{For TYC 2187-512-1, \cite{Quirrenbach2022} reported two $\rm [Fe/H]$ values: $0.08\pm0.19$ dex from \cite{Passegger2019} and $-0.18\pm0.08$ dex from \cite{Marfil2021}.}. We present the metallicity distributions of two groups in Figure~\ref{HK_comparison_cumulative} (HJ: red line, WJ: blue line). We examine whether the HJ and WJ populations are similar to each other through the same method as in Section~\ref{Giant_Planets_Favor_Metal_Rich_M_Dwarfs}. We find that the $p$-values of both the K-S and A-D tests are approximately $10^{-1}$. In fact, almost none of randomly drawn samples have $p$-values falling below 0.003. All results can be found in Table~\ref{fehksadtest}. Similar features are also seen for [M/H] (see Table~\ref{mhksadtest}). Consequently, we are not able to reject the null hypothesis that two distributions are identical. At this point, we conclude that HJs and WJs present analogous preferences on the metallicity of their host M dwarfs. As an independent inspection, we vary the $a/R_\ast$ boundary between 15 and 50 to define HJs and WJs and rerun the analysis. We find that the conclusion does not depend on the exact choice of $a/R_\ast$. 



\begin{table*}
    \centering
    \caption{The $p$-values of K-S and A-D tests of the $\rm [Fe/H]_{M13,H}$ (upper) as well as $\rm [Fe/H]_{M13,K}$ (lower) distributions. Four groups are investigated including field M stars, M dwarfs with giant planets along with its two subgroups: hot ($a/R_\ast\leq 20$) and warm ($a/R_\ast>20$) Jupiters. The value in the bracket is the fraction of randomly generated samples that have $p$-values smaller than 0.003 in our simulation (see Section~\ref{discussion}). All bottom left results are from the K-S test while the results on the top right are from the A-D test.}
    \label{fehksadtest}   
    \begin{tabular}{c|c|c|c|c}
        \hline\hline
        $\rm [Fe/H]_{M13,H}$ &Field M &M + HJ &M + WJ &M + HJ or WJ\\\hline
        Field M &$\cdots$ &$3.9\times 10^{-6}\ (99.2\%)$ &$1.9\times 10^{-6}\ (99.7\%)$ &$9.9\times 10^{-7}\ (100\%)$\\\hline
        M + HJ &$3.9\times 10^{-4}\ (83.2\%)$ &$\cdots$ &$4.4\times 10^{-1}\ (0\%)$ &$\cdots$\\\hline
        M + WJ &$1.4\times 10^{-4}\ (73.9\%)$ &$4.0\times 10^{-1}\ (0\%)$ &$\cdots$ &$\cdots$\\\hline
        M + HJ or WJ &$4.9\times 10^{-7}\ (100\%)$ &$\cdots$ &$\cdots$ &$\cdots$\\
        \hline
    \end{tabular}

    \begin{tabular}{c|c|c|c|c}
        \hline\hline
        $\rm [Fe/H]_{M13,K}$ &Field M &M + HJ &M + WJ &M + HJ or WJ\\\hline
        Field M &$\cdots$ &$4.1\times 10^{-4}\ (83.4\%)$ &$1.0\times 10^{-2}\ (49.2\%)$ &$4.2\times 10^{-5}\ (99.7\%)$ \\\hline
        M + HJ &$1.7\times 10^{-3}\ (60.1\%)$ &$\cdots$ &$1.4\times 10^{-1}\ (0\%)$ &$\cdots$\\\hline
        M + WJ &$1.6\times 10^{-2}\ (24.4\%)$ &$1.2\times 10^{-1}\ (0.1\%)$ &$\cdots$ &$\cdots$\\\hline
        M + HJ or WJ &$6.3\times 10^{-5}\ (95.2\%)$ &$\cdots$ &$\cdots$ &$\cdots$\\
        \hline
    \end{tabular}
\end{table*}

\begin{figure*}
\centering
\includegraphics[width=0.99\textwidth]{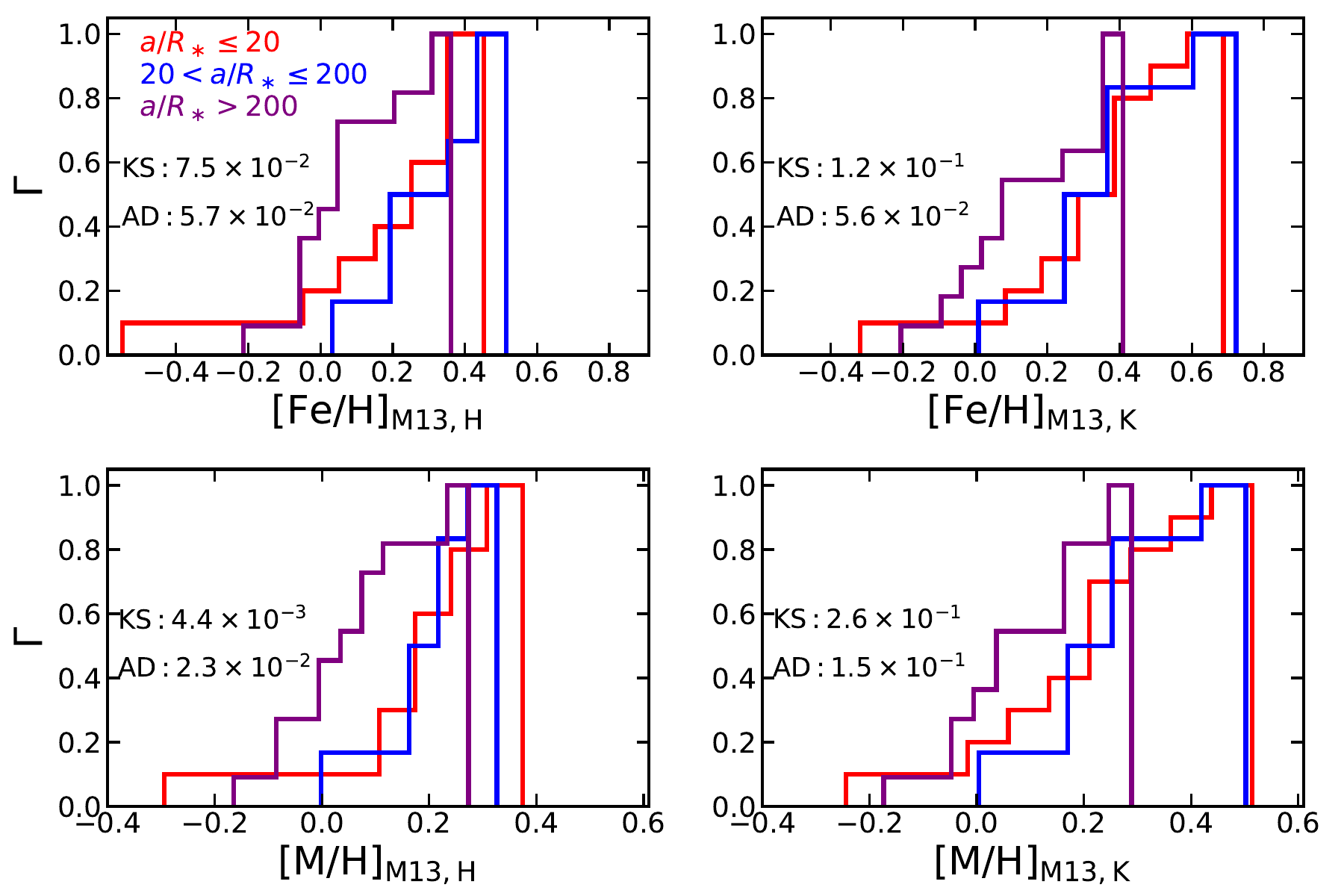}
\caption{The cumulative [Fe/H] (top) and [M/H] (bottom) distributions from two bands of M dwarfs hosting giant planets located in three different scaled semi-major axis bins: $a/R_\ast\leq 20$, $20<a/R_\ast\leq 200$ and $a/R_\ast>200$. The outer cold Jupiters with $a/R_\ast>200$ tend to show a weaker preference for metallicity compared with the inner hot Jupiters with $a/R_\ast \leq 20$. The numbers listed on the left of each panel are the $p$-values of the K-S and A-D tests between the HJ (red) and CJ (purple) populations.}
\label{hot_warm_cold}
\end{figure*}

Next, we attempt to explore the iron abundance distribution of cold Jupiters (CJs). The reason is that planets formed through gravitational instability are generally located at several AUs away from the host star \citep{Boss2006}. The massive gas giants could form around M dwarfs with relatively low metallicity in the context of such a framework \citep{Boss2002}. If the inner HJs and outer CJs are from two formation channels, we might expect to spot a discrepancy in their metallicity distribution. We further separate wide-orbit Jupiters with scaled semi-major axis $a/R_\ast$ larger than 200 from the original WJ sample, where we find 11 out of 17 planets belonging to this subgroup. The boundary of $a/R_\ast=200$ here is chosen somewhat arbitrarily by visually inspecting the $a/R_\ast$ distribution of our planet sample (see Figure~\ref{mp_a_rs} in the Appendix), where there is a blank of planets with $a/R_\ast$ between 200 and 400. Figure~\ref{hot_warm_cold} presents the cumulative [Fe/H] and [M/H] distributions of three categories: HJ ($a/R_\ast\leq 20$), WJ ($20<a/R_\ast\leq 200$) and CJ ($a/R_\ast> 200$). Under this new classification above, we find that the CJs tentatively show a weaker dependence on metallicity compared with HJs. The K-S and A-D tests show that the $p$-values vary between $4.4\times 10^{-3}$ and $2.6\times 10^{-1}$, shown in Figure~\ref{hot_warm_cold}. If such a phenomenon turns out to be true, wide-orbit ($a\gtrsim 0.5$~AU) gas giants around M dwarfs perhaps have an origin different from close-in ($a\lesssim 0.05$~AU) analogues. Given the limited sample size, however, we do not intend to claim any trend at present. More detections of CJs through long-term near-infrared spectroscopic surveys are required to distinguish the metallicity difference between the HJ and the CJ population.

\subsection{No Significant Correlation Between Stellar Metallicity and Planet Mass}

\begin{figure*}
\centering
\includegraphics[width=0.99\textwidth]{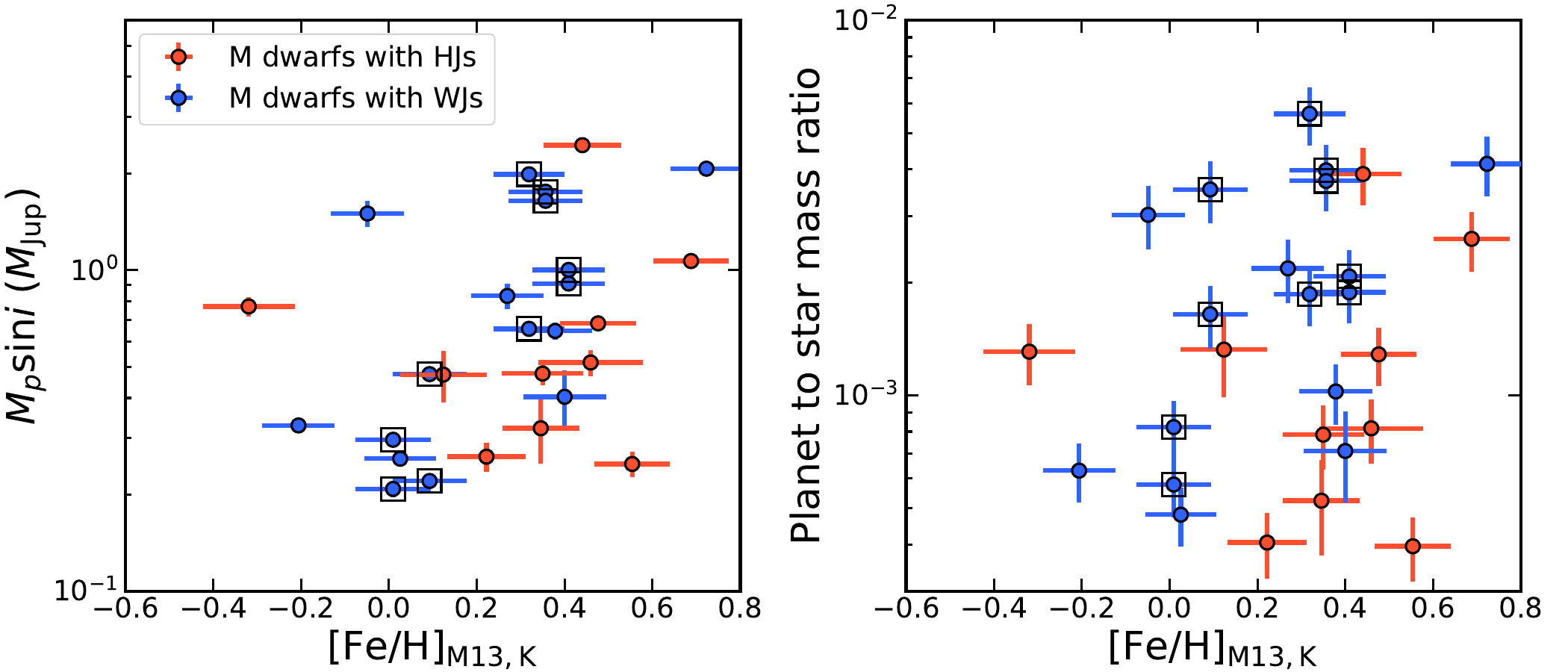}
\caption{\textit{Left panel:} Refined planet mass versus host metallicity $\rm [Fe/H]_{\rm M13,K}$. The red and blue dots are giant planets with $a/R_\ast\leq 20$ and $a/R_\ast >20$. Five multi-giant planet systems are marked with black boxes. \textit{Right panel:} Similar to the left but for mass ratio.}
\label{feh_mp}
\end{figure*}

Metal-rich stars are supposed to supply more solid materials supporting core accretion, and thus have the ability to form massive planets even around low-mass stars \citep{Ida2004}. In Figure~\ref{feh_mp}, we plot the iron abundances versus refined planet masses and the planet-to-star mass ratios. We compute the Pearson correlation coefficients between the mass $M_p$ of all giant planets in the planet sample and the $\rm [Fe/H]$ as well as $\rm [M/H]$ of their host stars, and we find the $p$-values have a large scatter, varying between 0.01 and 0.9. As all $p$-values are greater than 0.003, we consider there is no significant correlation between stellar metallicity and planet mass. Nevertheless, we note that there seems to be a lack of giant planets with mass above $1\ M_{\rm Jup}$ around low-metallicity M stars, similar to giant planets around Sun-like stars \citep[e.g.,][]{Fischer2005,Thorngren2016}. 

Additionally, massive hot Jupiters are rare around M dwarfs, manifesting as a lack of planets with $M_{p}\geq 1\ M_\mathrm{Jup}$ or $M_p/M_\ast \geq 2\times10^{-3}$. Moving outwards, radial velocity surveys have found several high-mass-ratio M dwarf-giant planet systems, probably due to observational bias. Multi-giant planets seem to always form far from the host M dwarfs, which have a similar [Fe/H] distribution as M stars hosting a HJ or a single WJ.

\subsection{Dependence on Stellar Mass}

Theoretical simulations show that mid-to-late M dwarfs are even more challenging to form giant planets compared with early-M stars \citep[e.g.,][]{Liu2019,Burn2021}, which might be compensated by a higher metallicity. We thus investigate whether stellar mass impacts the planet-metallicity relation.

\begin{figure}
\centering
\includegraphics[width=0.49\textwidth]{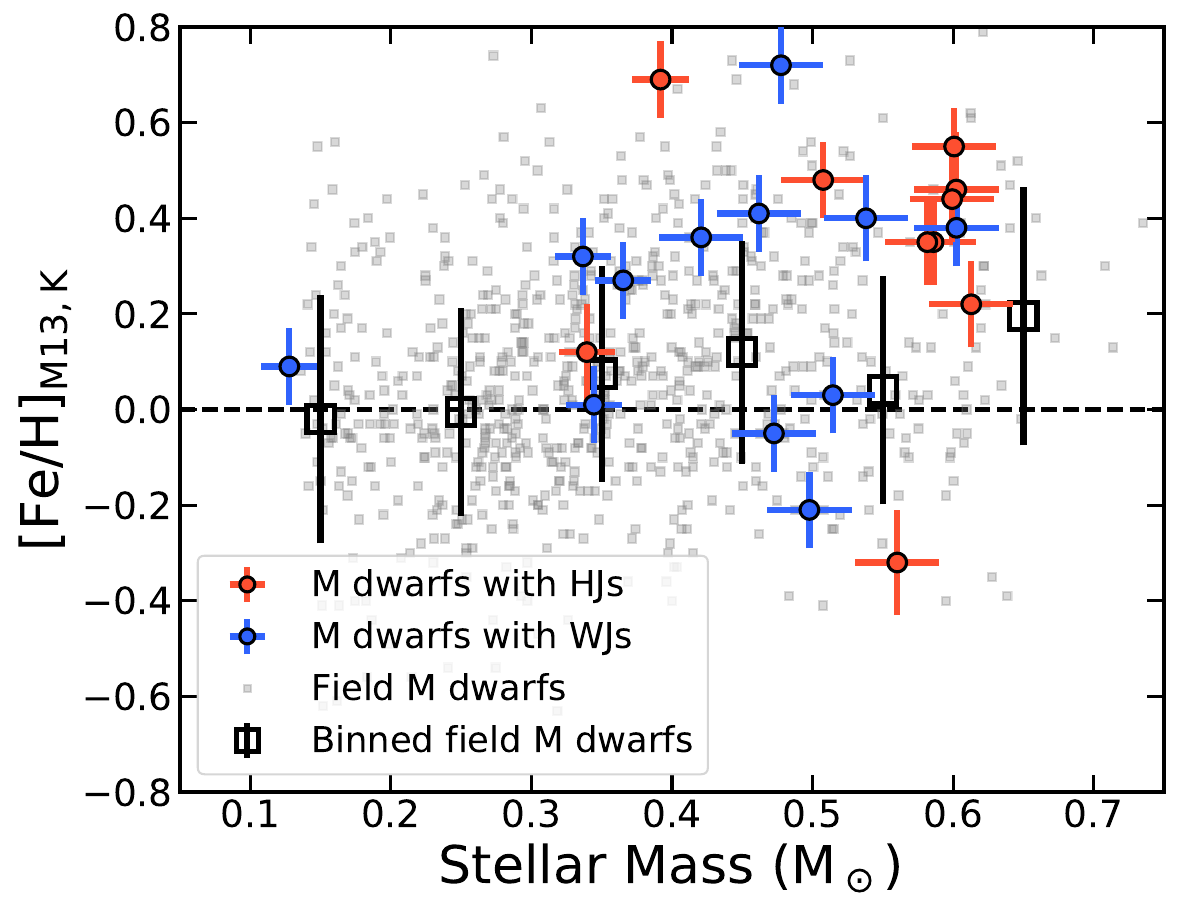}
\caption{The iron abundance $\rm [Fe/H]_{\rm M13,K}$ vs. stellar mass of 22 M dwarfs with HJs (red) and WJs (blue) in our planet sample. The background grey dots and black squares represent the field M dwarfs and their binned results (binning size=$0.1\ M_\odot$) where the uncertainties are the standard deviations in each stellar mass bin.}
\label{feh_ms}
\end{figure}

Figure~\ref{feh_ms} presents the iron abundance $\rm [Fe/H]_{\rm M13,K}$ and refined stellar mass of our M dwarf sample with confirmed giant planets as well as the field M star group. We find that the field M dwarfs mostly have a solar-like metallicity as their mass increases from 0.1 to 0.65 $M_\odot$. Early-M dwarfs ($M_\ast > 0.4\ M_\odot$) with giant planets have a wide range of metallicity, spanning from -0.4 to 0.7 dex. Both HJs and WJs have ever been detected around metal-poor early-M stars. Although mid-to-late M dwarfs ($M_\ast \leq 0.4\ M_\odot$) hosting giant planets have solar-like or super-solar metallicities, these values are not significantly higher than those of field stars within the same stellar mass range and those early-M dwarfs with gas giants, but only a few such systems have been confirmed. Future discoveries of such systems shall remedy the situation and enable a detailed comparison between giant planets around early and mid-to-late M dwarfs.

\section{Conclusions}\label{conclusion}

In this work, we present the homogeneously derived stellar and planet properties of 22 M dwarfs hosting 27 giant planets with spectra collected by IRTF/SpeX. By comparing the metallicity distribution of this planet sample with a field M dwarf group analyzed following the same procedure, we find that giant planets have a strong preference to metal-rich M dwarfs (4-5$\sigma$ significance), similar to FGK counterparts. Meanwhile, we find no evidence of metallicity dependence difference between hot Jupiters ($a/R_\ast\leq 20$) and warm Jupiters ($a/R_\ast> 20$) around M dwarfs. A subsample with $a/R_\ast>200$ that belongs to the warm-Jupiter group tends to have a weaker preference on both [Fe/H] and [M/H] compared with the hot Jupiter population. Based on the refined stellar and planetary physical parameters, we examine the stellar metallicities and planet masses and we find no significant correlation between them, which is yet to be confirmed given the limited sample size. Finally, M dwarfs with multi-giant planets and mid-to-late M dwarfs with giant planets do not show especially high metallicities compared with those hosting a single gas giant and those early-M dwarfs with giant planets. 


\section{Acknowledgments}
We thank the anonymous referees and editors for their comments that improved the quality of this publication. We thank Steven Giacalone and Emma Turtelboom for useful discussions. This work is supported by the National Science Foundation of China (Grant No. 12133005). T. G. acknowledges the Tsinghua Astrophysics High-Performance Computing platform at Tsinghua University for providing computational and data storage resources that have contributed to the research results reported within this manuscript. This work is based on data obtained with the NASA Infrared Telescope Facility, which is operated by the University of Hawaii. This research has made use of the NASA Exoplanet Archive, which is operated by the California Institute of Technology, under contract with the National Aeronautics and Space Administration under the Exoplanet Exploration Program. 

%

\vspace{5mm}
\facilities{IRTF/SpeX \citep{Rayner2003}}


\software{metal \citep{Mann2013}, SpeXtool \citep{Cushing2004}, SPLAT \citep{Burgasser2017}}



\appendix

\section{Results of overall metallicity [M/H]}

\begin{figure*}
\centering
\includegraphics[width=0.99\textwidth]{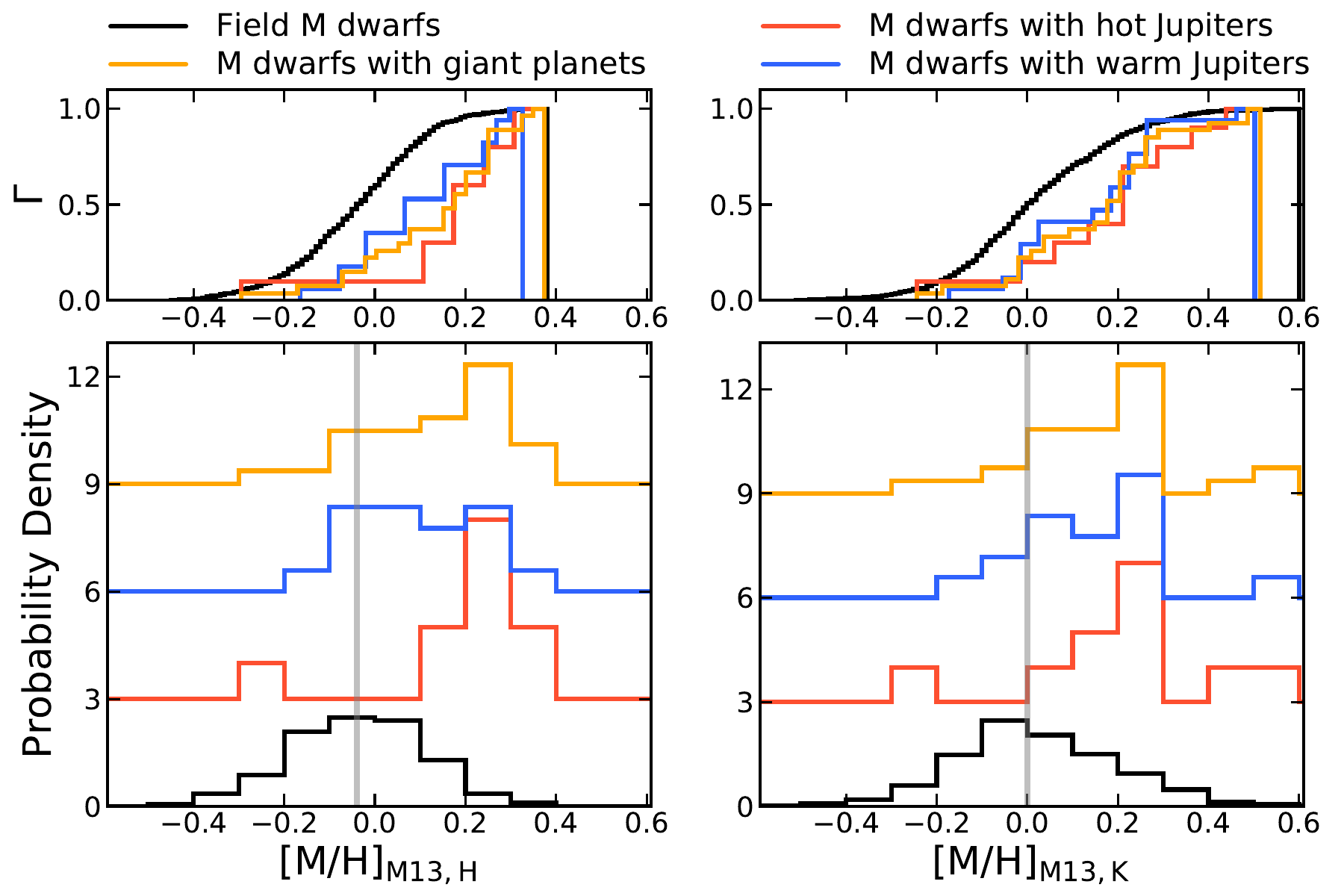}
\caption{Same as Figure~\ref{HK_comparison_cumulative} but for $\rm [M/H]_{M13,H}$ and $\rm [M/H]_{M13,K}$. The vertical grey line in each panel marks the median [M/H] of the field M dwarf sample, -0.04 and 0.0 dex. The results are similar to those using [Fe/H].}
\label{MH_HK_comparison_cumulative}
\end{figure*}

We rerun the analysis in Section~\ref{discussion} for all [M/H] measurements. The results are listed in Figure~\ref{MH_HK_comparison_cumulative} and Table~\ref{mhksadtest}. We find that the behavior of [M/H] is similar to those using [Fe/H]. 

\begin{table}
    \centering
    \caption{Same as Table~\ref{fehksadtest} but for $\rm [M/H]_{M13,H}$ and $\rm [M/H]_{M13,K}$. The results are similar to those using [Fe/H].}
    \label{mhksadtest}   
    \begin{tabular}{c|c|c|c|c}
        \hline\hline
        $\rm [M/H]_{M13,H}$ &Field M &M + HJ &M + WJ &M + HJ or WJ\\\hline
        Field M &$\cdots$ &$9.9\times10^{-7}\ (99.2\%)$ &$5.3\times10^{-5}\ (81.5\%)$ &$9.9\times10^{-7}\ (100\%)$\\\hline
        M + HJ &$5.1\times10^{-8}\ (93.9\%)$ &$\cdots$ &$1.0\times10^{-1}\ (0.7\%)$ &$\cdots$\\\hline
        M + WJ &$5.0\times10^{-3}\ (44.7\%)$ &$6.7\times10^{-2}\ (0.9\%)$ &$\cdots$ &$\cdots$\\\hline
        M + HJ or WJ &$3.6\times10^{-8}\ (99.6\%)$ &$\cdots$ &$\cdots$ &$\cdots$\\
        \hline
    \end{tabular}
    \\
    \begin{tabular}{c|c|c|c|c}
        \hline\hline
        $\rm [M/H]_{M13,K}$ &Field M &M + HJ &M + WJ &M + HJ or WJ\\\hline
        Field M &$\cdots$ &$1.1\times10^{-3}\ (50.9\%)$ &$2.8\times10^{-3}\ (25.0\%)$ &$1.5\times10^{-5}\ (98.3\%)$\\\hline
        M + HJ &$3.4\times10^{-3}\ (30.9\%)$ &$\cdots$ &$5.6\times10^{-1}\ (0\%)$ &$\cdots$\\\hline
        M + WJ &$6.6\times10^{-3}\ (13.3\%)$ &$7.6\times10^{-1}\ (0\%)$ &$\cdots$ &$\cdots$\\\hline
        M + HJ or WJ &$4.0\times10^{-5}\ (85.7\%)$ &$\cdots$ &$\cdots$ &$\cdots$\\
        \hline
    \end{tabular}
\end{table}

\section{Planet mass, scaled semi-major axis and equilibrium temperature distribution of the giant planet sample}

\begin{figure}
\centering
\includegraphics[width=0.99\textwidth]{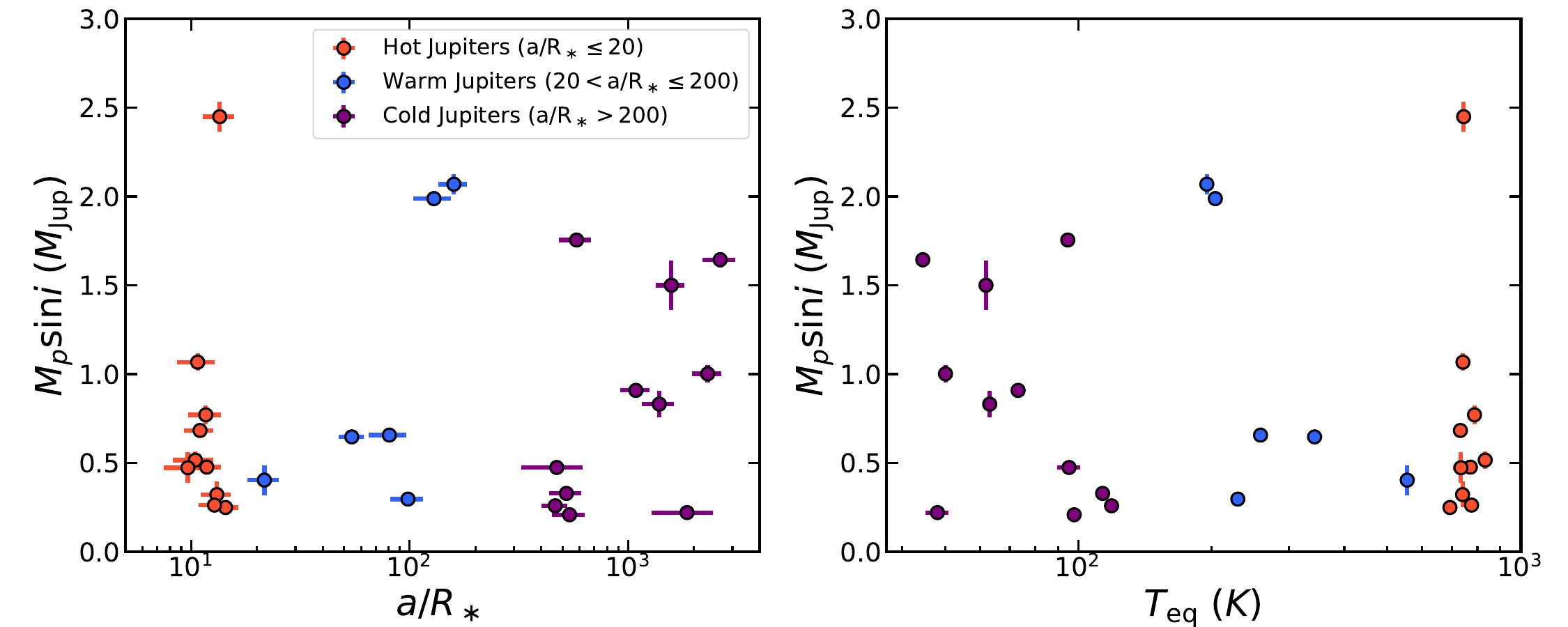}
\caption{{\it Left panel:} The planet mass and scaled semi-major axis distribution of our M dwarf giant planet sample. Different colors represent planets with scaled semi-major axis belonging to different ranges (see Section~\ref{Metallicity_Preference_of_Hot_and_Warm_Jupiters} for details). {\it Right panel:} Similar to the left but for equilibrium temperature.}
\label{mp_a_rs}
\end{figure}

In Figure~\ref{mp_a_rs}, we show the planet mass, scaled semi-major axis and equilibrium temperature distributions of our giant planet sample. We divide the full group into three categories based on their scaled semi-major axis: hot Jupiters ($a/R_\ast\leq 20$), warm Jupiters ($20<a/R_\ast\leq 200$) and cold Jupiters ($a/R_\ast> 200$). Such a designation is only applied when exploring the stellar metallicity difference between the hot Jupiter and cold Jupiter group (see Section~\ref{Metallicity_Preference_of_Hot_and_Warm_Jupiters}).

\bibliography{planet}{}
\bibliographystyle{aasjournal}



\end{document}